\documentclass[amsmath,superscriptaddress,preprint,aps]{revtex4-2}
\usepackage{graphicx}
\usepackage{dcolumn}
\usepackage{bm}
\usepackage{hyperref}
\usepackage{siunitx}
\usepackage{color, soul}

\setstcolor{red}
\usepackage[mathlines]{lineno}

\begin{document}
	
	\title{Van der Waals functionalization of ultrahigh-Q silica microcavities for $\chi^{(2)}$-$\chi^{(3)}$ hybrid nonlinear photonics}

	\author{Shun~Fujii}
	\email[Corresponding author. ]{shun.fujii@phys.keio.ac.jp}
 	\affiliation{Quantum Optoelectronics Research Team, RIKEN Center for Advanced Photonics, Saitama 351-0198, Japan}
 	\affiliation{Department of Physics, Faculty of Science and Technology, Keio University, Yokohama, 223-8522, Japan}

	\author{Nan~Fang}
 	\affiliation{Nanoscale Quantum Photonics Laboratory, RIKEN Cluster for Pioneering Research, Saitama 351-0198, Japan}
	
	\author{Daiki~Yamashita}
 	\affiliation{Quantum Optoelectronics Research Team, RIKEN Center for Advanced Photonics, Saitama 351-0198, Japan}
  \affiliation{Platform Photonics Research Center, National Institute of Advanced Industrial Science and Technology (AIST), Ibaraki 305-8568, Japan}
	
    \author{Daichi~Kozawa}
 	\affiliation{Quantum Optoelectronics Research Team, RIKEN Center for Advanced Photonics, Saitama 351-0198, Japan}
 	\affiliation{Nanoscale Quantum Photonics Laboratory, RIKEN Cluster for Pioneering Research, Saitama 351-0198, Japan}
	\affiliation{Research Center for Materials, National Institute for Materials Science, Ibaraki 305-0044, Japan}

     \author{Chee~Fai~Fong}
 	\affiliation{Nanoscale Quantum Photonics Laboratory, RIKEN Cluster for Pioneering Research, Saitama 351-0198, Japan}
 
    \author{Yuichiro~K.~Kato}
    \email[Corresponding author. ]{yuichiro.kato@riken.jp}
 	\affiliation{Quantum Optoelectronics Research Team, RIKEN Center for Advanced Photonics, Saitama 351-0198, Japan}
   	\affiliation{Nanoscale Quantum Photonics Laboratory, RIKEN Cluster for Pioneering Research, Saitama 351-0198, Japan}
	
	
\begin{abstract} 
Optical nonlinear processes are indispensable in a wide range of applications including ultrafast laser sources, microscopy, metrology, and quantum information technologies. Combinations of the diverse nonlinear processes should further lead to the development of unique functionalities, but simultaneous use of second- and third-order nonlinear processes is generally difficult. Second-order effects usually overwhelm the higher-order ones, except in centrosymmetric systems where the second-order susceptibility vanishes to allow the use of the third-order nonlinearity. Here we demonstrate a hybrid photonic platform whereby the balance between second- and third-order susceptibilities can be tuned flexibly. Ultrahigh-Q silica microcavities capable of generating third-order effects are functionalized by atomically thin tungsten diselenide, and we observe cavity-enhanced second-harmonic generation and sum-frequency generation with continuous-wave excitation at a power level of only a few hundred microwatts. Pump power dependence exhibits drastic increase and saturation of the second-harmonic light, originating from the dynamic phase-matching process. We show that the coexistence of second- and third-order nonlinearities in a single device can be achieved by carefully choosing the size and the location of the two-dimensional material. Our approach can be generalized to other types of cavities, unlocking the potential of hybrid systems with controlled nonlinear susceptibilities for novel applications.
\end{abstract}

	\maketitle
 
 
    \section*{Introduction}	
 Since the landmark discovery of second-harmonic generation (SHG)~\cite{Franken1961} enabled by the invention of lasers~\cite{MAIMAN1960}, nonlinear optics have played a central role in the development of diverse photonics applications. Frequency conversion processes are of particular importance, being extensively employed in ultrafast optics~\cite{Armstrong2004}, metrology~\cite{Reichert1999,Jones2000}, quantum state generation~\cite{Burnham1970,Slusher1985}, as well as microscopy~\cite{Squier1998,Shen1989}. To achieve these functionalities, both second- and third-order processes such as SHG, third-harmonic generation (THG), sum-frequency generation (SFG), parametric down conversion, and four-wave mixing (FWM) are utilized.

With such a variety of nonlinear effects, combinations of frequency conversion processes would allow for more flexible spectral synthesis. It is, however, generally difficult for second- and third-order processes to coexist. The efficiencies of nonlinear processes depend directly on the nonlinear susceptibility of conversion media but the origins are markedly different for second- and third-order susceptibilities. An essential requirement for second-order nonlinear processes to occur is inversion symmetry breaking, and typical materials include dielectric crystals (for example, lithium niobate and beta-barium borate), III-V semiconductors and organic crystals~\cite{Boyd2003}. Although third-order nonlinear susceptibility can simultaneously exist, second-order process dominates as higher-order nonlinearities are generally weak~\cite{Boyd2003}. Conversely, second-order nonlinear susceptibility vanishes in centrosymmetric crystals and amorphous materials (e.g., liquids, gases, and amorphous solids), and only third-order processes can be utilized in these $\chi^{(3)}$ materials. 

In this regard, one promising strategy is to establish a hybrid system by combining a non-centrosymmetric nonlinear material with an ultrahigh-Q microcavity fabricated from a $\chi^{(3)}$ material~\cite{Fryett2018}. The strength of second-order processes can be controlled through mode overlap with the non-centrosymmetric material, while exceptional enhancement of optical density in the tiny mode space can be facilitated to boost the third-order process to a practical level. 

As a candidate system, we propose ultrahigh-Q silica microcavities functionalized by transition metal dichalcogenides (TMDs). Silica whispering-gallery microcavities boast ultrahigh-Q properties ($>10^8$) that ensure high-circulating optical intensities essential to induce various third-order optical nonlinear processes~\cite{Kippenberg2004,Kippenberg2004a,DelHaye2013,Farnesi2014,Yi2015,ChenJinnai2016,Fujii2017}. Meanwhile, monolayer TMDs possess a comparable magnitude of second-order nonlinearity to commonly used nonlinear crystals~\cite{Malard2013,Woodward2016,Autere2018} and are thus expected to be used for practical nonlinear applications~\cite{Lin2019,Klimmer2021,Trovatello2021,Xia2014,Ngo2022}. Their atomically thin nature gives them mechanical flexibility to conform to the surface of the optical microcavities and the van der Waals character makes them compatible for the heterogeneous interface~\cite{Javerzac-Galy2018,Tan2021,He2021,Nan2023}.


Here, we demonstrate a novel nonlinear photonic platform by functionalizing ultrahigh-Q silica microspheres with tungsten diselenide ($\mathrm{WSe_2}$). Atomically thin layers of the two-dimensional (2D) material are transferred onto the cavity with a minimal level of scattering loss.  Cavity-enhanced second-harmonic (SH) generation is achieved by a continuous-wave (CW) excitation with only a few hundreds of microwatts thanks to strong light-matter interaction between a resonant optical field and integrated $\mathrm{WSe_2}$. We also observe efficient SFG with a two-color excitation scheme. In addition, pump power dependence shows self-locking of SH output, revealing the mechanism of the dynamic phase-matching process. It is confirmed that the SH process only occurs for odd layer numbers, and the coexistence of second- and third-order nonlinearities in a single device is achieved by controlling the second-order susceptibility of the device. 




	 \begin{figure*}
	\centering
		\includegraphics[width=\textwidth]{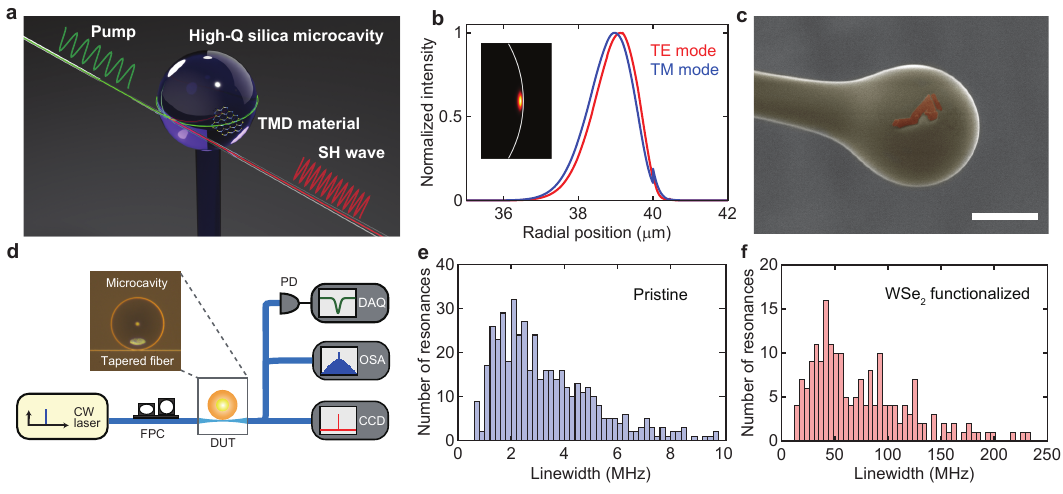}
		\caption{\label{Fig_concept} Functionalization of a high-Q silica microcavity by atomically thin 2D material. (a) Conceptual illustration of a monolayer-material-integrated silica microcavity realizing strong light-matter interaction. (b) Simulated normalized intensity of the optical mode across the equator of  a silica microsphere with a radius of 40~\textmu m. The calculations are conducted by using a finite element method (FEM) software (COMSOL Multiphysics). The cavity modes exhibit a slight difference in the profiles, and TM modes exhibit evanescent fields slightly higher than TE modes. The inset shows the optical mode profile, where the white line indicates the boundary between the silica and surrounding air. (c) False-color scanning electron micrograph image of a $\mathrm{WSe_2}$ integrated high-Q microsphere. The scale bar represents 50~\textmu m. (d) Experimental setup. FPC, fiber polarization controller; DUT, device under test; PD, photodetector; DAQ, data acquisition; OSA, optical spectrum analyzer; CCD, charged-coupled device installed to a spectrometer. (e, f) Histogram of cavity linewidths in pristine and functionalized microcavities. The degradation of Q-factors is mainly attributed to an increase in surface scattering loss.}
	\end{figure*}
	
\section*{Results and Discussion}
 \begin{figure*}
	\centering
		\includegraphics[width=\textwidth]{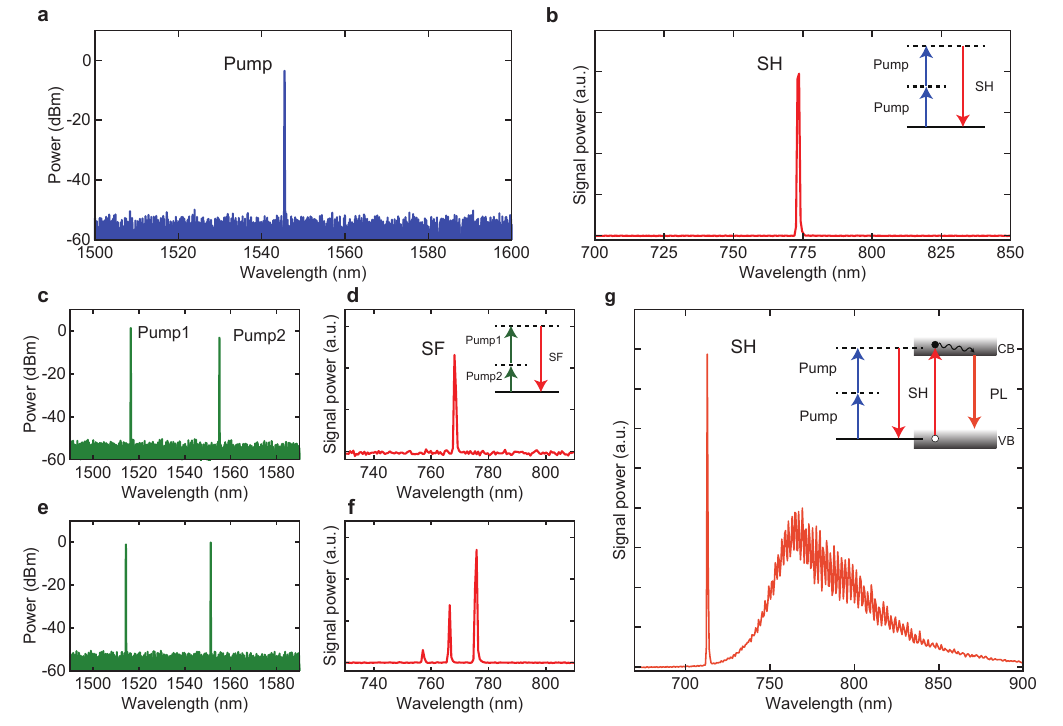}
		\caption{\label{Fig_shg_spectrum} Observation of second-order nonlinear processes in  material integrated microresonators. (a, b) Optical spectra of pump wavelength and generated SH light. The frequency of the SH light exactly matches twice the frequency of the pump light, indicating the nonlinear frequency-doubling process. The energy diagram of an SH process is shown in the inset. (c, d) Optical spectra of two different pump wavelengths and generated second-order sum-frequency (SF) light. The frequency of SF light corresponds to the sum of pump frequencies as depicted in the inset. (e, f) Measured spectra of pump wavelengths and corresponding visible light, where the two-color excitation scheme enables simultaneous generation of SFG and SHG. The difference in signal powers is owing to the phase-matching condition for near-infrared and visible cavity modes. (g) Optical spectra of SHG-mediated photoluminescence (PL) emission in a $\mathrm{WSe_2}$-functionalized cavity. The SH light at a wavelength of 715~nm excites excitonic PL in a monolayer $\mathrm{WSe_2}$, where the broad emission is optically coupled to numerous cavity modes. The inset shows the energy diagram of the process. CB: conduction band, VB: valence band.}
	\end{figure*}
 
Figure~\ref{Fig_concept}(a) shows a conceptual illustration of a 2D-material functionalized silica microcavity, capable of serving as a second-order nonlinear photonic platform. Strong light-matter interaction assisted by cavity resonance permits efficient nonlinear optical processes that originate from the atomically thin layered material with low-power CW excitation. The frequency converted light that resonates with another longitudinal resonance mode, in a situation referred to as a doubly-resonant condition, allows the cavity-enhanced signals to couple into the same waveguide coupler utilized for excitation. 

The normalized mode intensity of a microsphere cavity is shown in Fig.~\ref{Fig_concept}(b), where the inset shows the optical mode profile. Although the optical density is maximized at a few microns inside from the surface of the cavity, evanescent field exists at the boundary, which interacts with a surface integrated 2D material. These simulation results indicate that the evanescent tails, accounting for up to a few percentages of the total light intensity, can interact efficiently with the surface layer for both primary polarization modes, namely the transverse-electric (TE) and transverse-magnetic (TM) modes. The light-matter interaction on the surface layer generally increases in a smaller cavity because the ratio of the evanescent field to the total intensity is inversely related to the cavity radius. (The relationship between the evanescent field ratio and the cavity radius is further detailed in Supplementary Note 1.)

We first functionalize a silica microsphere cavity (diameter $\sim$80~\textmu m) by transferring mechanically exfoliated monolayer $\mathrm{WSe_2}$ onto the cavity surface using the polydimethylsiloxane (PDMS)-assisted dry-transfer technique~\cite{CastellanosGomez2014}. The layer number of $\mathrm{WSe_2}$ flakes are identified either through photoluminescence (PL) measurement~\cite{Zhao2013} or by optical contrast in microscope images prior to the transfer~\cite{Li2013rapid}. Figure~\ref{Fig_concept}(c) shows a false-color image of the $\mathrm{WSe_2}$ functionalized silica microsphere cavity. (The details on the sample fabrication are presented in Methods and Supplementary Note 2.) 

To characterize the influence of $\mathrm{WSe_2}$ flake on the Q-factor of a microcavity, we compare the transmission spectra before and after the transfer process. The experimental setup is presented in Fig.~\ref{Fig_concept}(d). All resonances observed within 1530 to 1570~nm are numerically fitted to a Lorentzian function. This allows for the statistical analysis of loaded FWHM linewidth (=$\omega/Q$) as shown in Figs.~\ref{Fig_concept}(e) and \ref{Fig_concept}(f). The median value in a pristine (i.e., before transfer) microsphere is 2~MHz, which corresponds to an ultrahigh Q-factor of $1\times10^8$. After the transfer of a $\mathrm{WSe_2}$ flake, the most probable loaded linewidth broadens to approximately 40~MHz, corresponding to a Q-factor of $5\times10^6$ even though the highest Q values remain about $10^7$.  

The degradation in the Q-factor is likely due to an increase in scattering loss resulting from the functionalization as is also observed in the integration of  materials to other nanophotonic cavities~\cite{Wu2015,Fryett2016,Javerzac-Galy2018}. We anticipate minimal effect on the Q-factor from the absorption loss caused by the $\mathrm{WSe_2}$ flake  because the telecom band photon energy is significantly lower than the bandgap of monolayer $\mathrm{WSe_2}$ ($\sim$1.75~eV). It should be noted that the uniformity of transferred flakes is the key to maintaining high Q-factors as well as the flake size and the transfered position, and placing a small flake away from the equator of a microcavity would greatly reduce the scattering loss in high-Q modes. For most of this study, however, we place priority on using uniform and large $\mathrm{WSe_2}$ flakes and transfer onto the equator of the device to maximize the interaction length between the optical modes and the $\mathrm{WSe_2}$ material.

	 \begin{figure*}
	\centering
		\includegraphics[width=\textwidth]{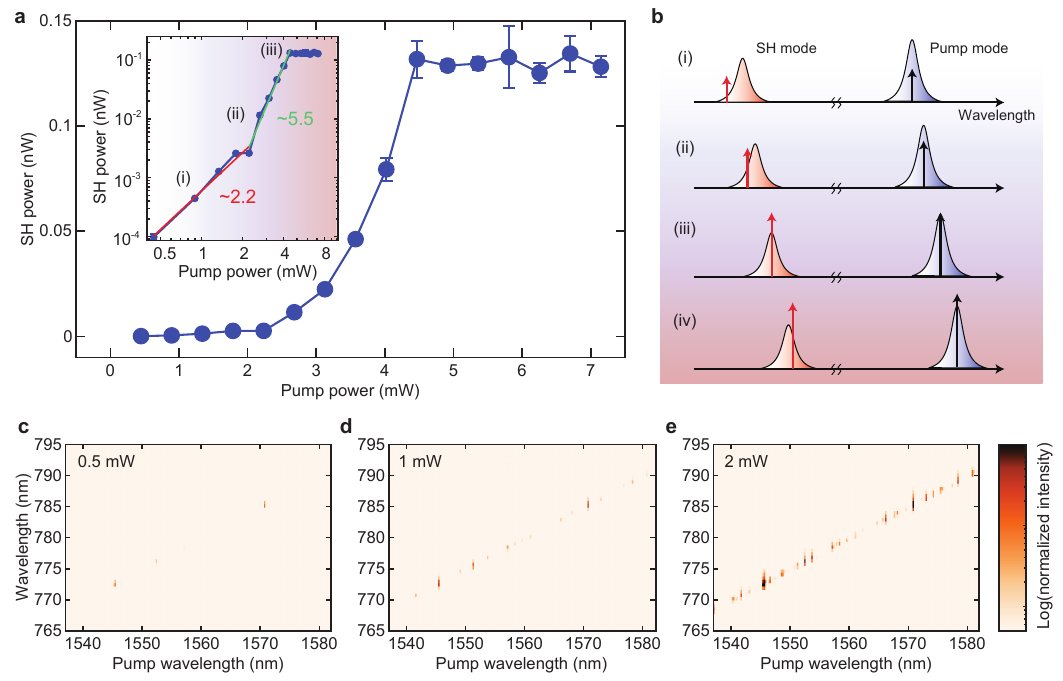}
		\caption{\label{Fig_shg_power} SH power dependence on the pump power. (a) Maximum SH power as a function of pump power for the same cavity mode. The data are presented in a log-log scaling with power-law fits in the inset. The SH powers exhibit a slope of $\sim$2.2 at the relatively low pump power regime ($<2$~mW), but the slope increases to $\sim$5.5 at the intermediate region (2--4.5~mW). At the high pump power region ($>$4.5~mW), the SH power starts to saturate and remains stable. The error bars correspond to the standard deviation of ten repeated measurements. (b) Schematic for the mechanism of dynamic phase-matching process. (c-e) Spectral mapping for different pump powers. The SH power exhibits significant dependence on the pump powers of 0.5~mW, 1~mW, and 2~mW.  The colormaps are normalized to a common scale.}
	\end{figure*}

	 \begin{figure*}
	\centering
		\includegraphics[width=\textwidth]{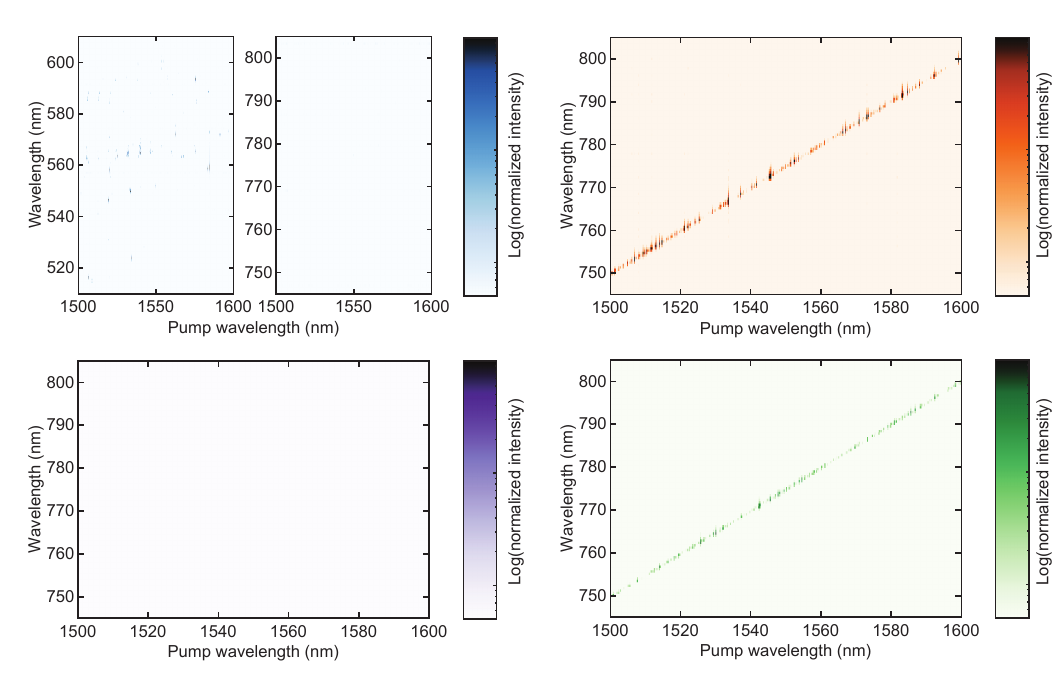}
		\caption{\label{Fig_shg_map} Layer dependence of SH light intensity. (a) Spectral mapping of signal intensity in the visible wavelength region in a pristine silica microsphere. Third-order nonlinear processes (e.g., THG, TSFG) are observed in the 500--600~nm wavelength range, whereas no strong signal appears in the 740--810~nm wavelength range due to the absence of the second-order nonlinearity. (b) Spectral mapping in a monolayer $\mathrm{WSe_2}$ functionalized microcavity. The strong SH signals are observed over a wide range of pump wavelengths (1500--1600~nm). (c) Mapping in a 2L-$\mathrm{WSe_2}$ functionalized sample. No clear SH light is measured in the map because the inversion symmetry exists in a 2H-stacked bilayer $\mathrm{WSe_2}$, thus the $\chi^{(2)}$ nonlinearity vanishes. (d) Mapping in a 3L-$\mathrm{WSe_2}$ device. The distinct SH light is observed again similar to the monolayer case since the inversion symmetry is broken for 3L-$\mathrm{WSe_2}$ crystal structures.  The measurements are performed with a fixed pump power of 3~mW.}
	\end{figure*}

 Efficient nonlinear optical processes are achieved by dynamically obtaining resonant phase-matching from thermal and Kerr nonlinearities. The phase-matching condition described here is different from the conventional phase-matching for frequency conversion processes in nonlinear crystals, which is usually satisfied by aligning the incident angle and polarization of the pump lasers. The situation becomes more complicated in a microcavity system due to resonance effects as well as inherent material and geometric cavity dispersion~\cite{Fujii2020}. Mode mismatch at the SH frequency is induced by the dispersion which hinders perfectly resonant phase-matching condition. In order to locate resonances with efficient frequency conversion, we carefully scan the pump wavelength while monitoring the visible spectrum with a thermoelectrically cooled charged coupled device (CCD) attached to a spectrometer. Continuous wavelength scan of the pump laser induces changes in the refractive index of the cavity via thermal and Kerr effects, thus giving rise to the redshift of the resonance modes. Since cavity modes with different resonant frequencies and spatial mode distribution experience different amounts of the redshift, the phase-matching condition can be fulfilled at certain pump wavelengths where efficient frequency conversion would occur.

 Figure~\ref{Fig_shg_spectrum} presents optical spectra in the visible and the corresponding pump wavelength bands. By carefully tuning the pump laser wavelength to a cavity resonance with a pump power of 500~\textmu W, second-harmonic (SH) light is clearly observed (Fig.~\ref{Fig_shg_spectrum}(a) and \ref{Fig_shg_spectrum}(b)). The frequency of the SH light (773.1~nm) exactly matches twice the pump frequency (1545.5~nm) with a wavelength error of only 0.045\%, and this fact confirms the occurrence of frequency doubling process via second-order optical nonlinearity. We stress that other third-order (Kerr) nonlinear processes, which could arise from bulk silica microcavities, are absent in this experiment because the threshold powers are far beyond our pump power level. The required pump powers for FWM and Raman oscillation are 12.6~mW and 36.1~mW, respectively, in the case of a loaded Q-factor of $5\times10^6$, as threshold powers of these processes scale as $V/Q^2$~\cite{Kippenberg2004a,Kippenberg2004}. (See Methods for details in theoretical estimation of the threshold powers). 
 
 Next,  we pump the device by using two CW lasers with different frequencies (i.e., two-color excitation) at sub-milliwatt pump powers. This scheme allows us to observe SFG as shown in Figs.~\ref{Fig_shg_spectrum}(c) and \ref{Fig_shg_spectrum}(d).  Two-color pump imposes a triply resonant condition on the sum-frequency process to be phase-matched, but it is easy to find the phase-matching condition by slowly tuning one laser while keeping the frequency of the other laser within a high-Q resonance. Figures~\ref{Fig_shg_spectrum}(e) and \ref{Fig_shg_spectrum}(f) show a unique example, where two SH and one SF light are generated from two laser input because of five-fold resonant triple phase-matching. 

In addition to second-order nonlinearities, we also observe excitonic photoluminescence (PL) from the monolayer $\mathrm{WSe_2}$. Figure~\ref{Fig_shg_spectrum}(g) shows a spectrum of SHG at a wavelength of 715~nm and the associated PL emission when pumping the device at a wavelength of 1530~nm. The multiple spikes seen in the PL spectrum indicate that broad excitonic PL couples to the high-Q cavity modes and the intensities are enhanced due to Purcell effect or modulated by the differing collection efficiencies. The energy diagram is depicted in the inset of Fig.~\ref{Fig_shg_spectrum}(g). We emphasize that the observation of this unique resonance energy transfer, i.e, SHG-mediated PL and subsequent resonant enhancement, has only become possible by our $\mathrm{WSe_2}$-functionalized high-Q devices. This result also proves the strong interaction between a monolayer $\mathrm{WSe_2}$ and whispering-gallery modes via an evanescent field.


The dynamic phase-matching mentioned previously is highlighted in the pump power dependence of the SH power as shown in Fig.~\ref{Fig_shg_power}(a). We measure the SH power for the same cavity mode, and carefully tune the pump wavelength so that the SH light is maximized at each pump power. This measurement scheme allows us to find the perfect phase-matching condition at a certain pump power, which can be dynamically altered by the nonlinear resonance shifts. The double logarithm plot is presented in the inset of Fig.~\ref{Fig_shg_power}(a), where three distinct regimes can be recognized. Below a pump power of $\sim$2~mW, the SH powers exhibit a linear slope of $\sim$2.2, which is very close to the anticipated slope of 2 for a SHG process. As the pump power is increased from 2~mW to 4.5~mW, the fitted slope drastically changes to $\sim$5.5, and further increase in the pump power (above 4.5~mW) induces saturation of the SH power. Such a kink behavior of the SHG intensity has not been reported in conventional SHG measurements of TMD flakes on substrates~\cite{Li2013,Kumar2013} or photonic nanostructures~\cite{Fryett2016,Liu2016,Chen2017,Ngo2022}. 

We therefore consider the influence of dynamic phase-matching condition in a double resonance system. Figure~\ref{Fig_shg_power}(b) shows the schematic for the mechanism under consideration, where the SH light is blue-detuned at low pump powers. In this scenario, SH light is considered to be almost in an off-resonance condition with a large detuning (state (i)), yielding a moderate conversion efficiency with a slope of approximately 2.  As the pump power increases, thermal and Kerr nonlinearities induce significant redshift of the resonances~\cite{Carmon2004}. While the frequency of the SH light  is twice the pump frequency (i.e., $\omega_{\mathrm{SH}}=2 \omega_{\mathrm{p}} $), the resonance mode $\omega_2$ for SH generally shows smaller shifts than the SH light ($\Delta \omega_2<\Delta \omega_{\mathrm{SH}} $) due to the imperfect mode overlap between the pump and SH modes~\cite{Zhang2019}. The detuning of the SH light therefore decreases at a higher pump power, leading to a rapid increase in conversion efficiency (state (ii)). Once the SH power reaches its maximum when  both cavity modes exactly match the on-resonance condition (state (iii)), a further increase in intracavity power results in the red detuning of SH light which would reduce the output (state (iv)). The maximum SH power for higher pump powers would then be obtained for the specific intracavity power where the double resonance condition is retained. Since the intracavity power is almost constant, the SH power saturates despite a further increase in the pump power. Such a complex power dependency is clearly observed in a separate experiment, where we record SH signals while continuously scanning the pump laser frequency at a certain pump power. As shown in Figs.~\ref{Fig_shg_power}(c)-\ref{Fig_shg_power}(e), the SH signal becomes more and more frequent in the spectral map, and the intensity is drastically enhanced with the increase in the pump power. We note that no pump polarization dependence is observed. (Extended data are presented in Supplementary Note 3.)

It is possible to calculate the conversion efficiency from the data in Fig.~\ref{Fig_shg_power}(a). When we define $P_{\mathrm{SH}}$ as the detected SH power, the calculated maximum conversion efficiency $P_{\mathrm{SH}}/P_\mathrm{p}^2$ is $6.6\times10^{-4}$~\%W$^{-1}$ with the pump power $P_\mathrm{p}$ of 4.5~mW. It should be noted that the internal (intracavity) conversion efficiency is expected to be over one order of magnitude higher than the above value because the waist of the nanofiber waveguide is optimized to the pump wavelength band in this experiment, thus resulting in poor coupling efficiency of SH light due to the phase mismatch between the visible band and the nanofiber coupler~\cite{Humphrey2007,Jiang2017}. We note that the collection efficiency can be improved by employing an additional nanofiber designed for SH wavelengths, i.e., add-drop configuration~\cite{Zhang2019,ChenJinnai2016} or by exploiting a chaotic channel in deformed microcavities~\cite{Jiang2017}.


As mentioned earlier, symmetry plays an important role in determining the nonlinear susceptibility, and therefore the number of layers in the two-dimensional material is a crucial factor. The $\mathrm{WSe_2}$ crystals used in this work possess the 2H-phase (semiconducting) structure, which is a more stable form than other crystal phases. The 2H-phase TMD crystals belonging to $D_{3h}$ space group exhibit substantial second-order nonlinearity only for odd layer numbers, whereas the $\chi^{(2)}$ nonlinearity vanishes in even layer numbers since the net nonlinear dipoles are cancelled out due to inversion symmetry~\cite{Li2013,Kumar2013}. Considering these selection rules, we perform a comparative experiment in four different devices: pristine silica device, ML-$\mathrm{WSe_2}$ functionalized device, 2L-$\mathrm{WSe_2}$ functionalized device, and 3L-$\mathrm{WSe_2}$ functionalized device. 

Figure~\ref{Fig_shg_map} shows the mapping of SH spectra in the visible wavelength region when the pump wavelength is scanned from 1500~nm to 1600~nm with a pump power of 3~mW. As we anticipate, strong SH light appears only in the ML and 3L-$\mathrm{WSe_2}$ devices (Figs.~\ref{Fig_shg_map}(b) and \ref{Fig_shg_map}(d)), whereas there is no distinct signal in the pristine and 2L-$\mathrm{WSe_2}$ devices (Figs.~\ref{Fig_shg_map}(a) and \ref{Fig_shg_map}(c)). This is clear evidence that second-order nonlinearity originates from the integrated $\mathrm{WSe_2}$, not from intrinsic surface symmetry breaking of the cavity material~\cite{Zhang2019}. In a pristine device, third-order processes such as THG and third-order SFG associated with pump, FWM, and stimulated Raman scattering (SRS) are observed in the range of 500--620~nm (Fig.~\ref{Fig_shg_map}(a), left) thanks to unaltered ultrahigh-Q properties ($>5\times10^7$). We find that the number of SH signal peaks in the map is surprisingly high in both ML- and 3L-$\mathrm{WSe_2}$ devices even though the Q-factors of most resonances are not as high as $10^7$. We attribute the efficient, highly populated SHG to giant second-order nonlinearity of TMD materials and relaxed resonant phase-matching condition due to cavity linewidth broadening. If we could achieve much higher Q-factors with larger overlap between the cavity mode and the  material, the conversion efficiency is expected to substantially increase; nevertheless, the resonant phase-matching condition would become stricter as a trade-off. 
 
 	 \begin{figure}
	\centering
		\includegraphics[]{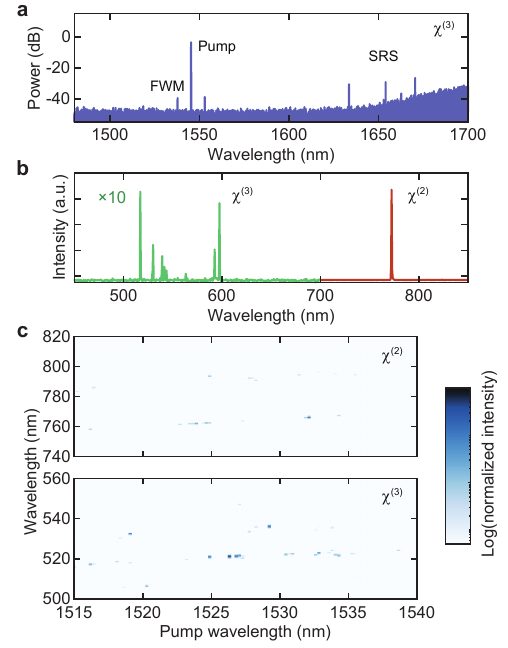}		\caption{\label{Fig_shg_coex} Coexistence of second- and third-order nonlinearities. (a, b) Optical spectra of telecom and visible wavelength regions, respectively. The frequency-converted light is generated via $\chi^{(3)}$ nonlinearity in the pump band, resulting in the complex spectrum in the visible band. (c) Spectral mapping of signal intensities in the two different wavelength bands, from 740~nm to 820~nm and from 500~nm to 560~nm. The distinct signals are observed in both bands, revealing the coexistence of second- and third-order nonlinearities.}
	\end{figure}
 
We have shown thus far the results focused on the emergence of second-order nonlinearity, but one key advantage of this technique is its flexible controllability of nonlinear susceptibility. By carefully controlling the transferred position and the flake size of  materials, it is possible to tune the balance between second- and third-order nonlinearity. Here, we intentionally place a small flake away from the equator of a cavity to keep Q-factors high enough ($>10^7$) to simultaneously observe both second- and third-order nonlinear processes in the same device.

 Figures~\ref{Fig_shg_coex}(a) and \ref{Fig_shg_coex}(b) show the observed optical spectra in the pump and the visible wavelength bands in this $\mathrm{WSe_2}$ functionalized microcavity. In the pump wavelength band, FWM sidebands are observed in the vicinity of the pump light and a few Raman peaks can be recognized around 1630--1670~nm, which coincides with the Raman gain band of silica~\cite{Kippenberg2004a}. For the visible wavelength band, the peaks around 520--600~nm arise from THG and third-order SFG processes involving the peaks seen in the pump band. The signals around 600~nm are believed to involve a cascaded Raman process~\cite{Farnesi2014,ChenJinnai2016,Fujii2017}. While these signals originate from third-order nonlinearity, the strong signal at a wavelength of 772~nm corresponds to the SH light of the pump light via second-order nonlinearity induced by monolayer $\mathrm{WSe_2}$. The spectral map is shown in Fig.~\ref{Fig_shg_coex}(c), where the strong visible light is recognized as a result of simultaneous generation of second- and third-order processes. The signals around 780--800~nm come from the second-order SFG process of the pump and Raman components, which are not observed in the previous experiments (Fig.~\ref{Fig_shg_map}). 

\section*{Conclusions}
In conclusion, we have demonstrated a novel approach for introducing second-order optical nonlinearity in ultrahigh-Q silica microcavities through functionalization by a two-dimensional material. By integrating atomically thin TMD layers with broken crystal inversion symmetry onto the surface of amorphous silica microspheres, cavity-enhanced SHG and SFG arise from strong light-matter interaction via evanescent fields. The cavity-enhanced PL emission mediated by the SHG process reveals the distinct optical coupling between SH light and the excitonic resonance of the monolayer $\mathrm{WSe_2}$. The conversion efficiency of SH light is strongly dependent on the pump power as a result of the dynamic phase-matching process, leading to the drastic increase and saturation of the SH power. A carefully coordinated clean-stamp transfer technique allows for investigation of layer number dependence as well as manipulation of the relative strength of the second- and third-order optical nonlinearity in the device.

Practical levels of second-order nonlinearity in $\chi^{(3)}$ materials has long been strongly desired. Surface symmetry breaking~\cite{Zhang2019,Levy2011} and photo-induced effects~\cite{Lu2021,Nitiss2022} can introduce second-order nonlinear susceptibility but are limited in various aspects. In comparison, this study offers a powerful way to controllably enhance optical nonlinearity in high-Q microcavities, which would bring breakthroughs in nonlinear optics. The results presented in this work lead to an anticipation that optical nonlinearity can be artificially designed in hybrid systems where various nonlinear processes are combined to implement unconventional functionalities.


In addition, we note that this approach can be extended to other centrosymmetric high-Q cavity devices, including integrated ring resonators made of silicon or silicon nitride ($\mathrm{Si_3 N_4}$), and thus paves the way to few-photon coherent nonlinear optics and quantum photon manipulation in various platforms. The combination of ultrahigh-Q cavities with nanomaterials opens up a novel regime in investigation of optical processes at high fields under CW excitation, potentially leading to intriguing physical phenomena as well as nanophotonic applications.

\section*{Methods}
\subsection*{Device fabrication}
A silica microsphere cavity is fabricated from a standard single-mode fiber (SMF-28) via arc-discharging using a commercial fiber fusion splicer. The cavity size can be roughly controlled by the discharge power, position, and duration. The arc-melting process ensures the surface uniformity and smoothness, enabling ultrahigh Q-factors of up to $10^8$. Before the cavity fabrication, we reduce the fiber diameter to approximately 20--30~\textmu m by adiabatically stretching the optical fiber. This preprocess allows the fabrication of a microsphere with a diameter of less than 100~\textmu m. 
Atomically thin $\mathrm{WSe_2}$ flakes are prepared by mechanical exfoliation of a bulk $\mathrm{WSe_2}$ crystal (HQ Graphene). The $\mathrm{WSe_2}$ flake is deposited on the microsphere surface using a dry-transfer technique assisted by a PDMS sheet. Precise position alignment is accomplished by using a motorized position system and a high-magnification microscope.

\subsection*{Experimental setup details}
A wavelength-tunable CW laser (Santec, TSL-710) at the telecom band is used as a pump light source. The polarization of the input light is adjusted to the desired cavity modes by using a fiber polarization controller (FPC). A tapered fiber waveguide (diameter of $\sim$1 \textmu m) is used to couple the pump light to a cavity via evanescent coupling.  The distance between the fiber and the cavity is carefully controlled by using a high-precision positioner since the coupling strength is determined by the distance and the coupling position~\cite{Humphrey2007,Fujii2023}. The pump (telecom band) and visible (400--900~nm) spectra are measured by an optical spectrum analyzer (OSA) and a thermoelectrically-cooled charge coupled device (CCD) camera attached to a high-resolution spectrometer, respectively. The transmission spectrum is recorded by using a photodetector (PD) and a data acquisition (DAQ) system triggered with laser wavelength sweeping. The laser power is kept less than 10~\textmu W to avoid thermal broadening of a resonance for the transmission measurement. The spectral mapping is recorded with a spectrometer while slowly scanning the pump laser wavelength.


\subsection*{Theoretical estimation of threshold power for third-order nonlinear processes}
 Threshold pump power for parametric FWM and Raman oscillation is obtained by taking into account the power build-up factor in a microcavity coupled to an external waveguide, respectively~\cite{Kippenberg2004,Kippenberg2004a},

 \begin{equation}
     P^\mathrm{FWM}_\mathrm{in} = \frac{\omega n^2 V_\mathrm{eff}}{8 Q^2 n_2 c \eta},
 \end{equation}
  \begin{equation}
     P^\mathrm{SRS}_\mathrm{in} = \frac{\omega^2 n^2 V_\mathrm{eff}}{4 Q^2 c^2 \eta g_R},
 \end{equation}
    where $n$ is the refractive index, $V_\mathrm{eff}$ is the effective mode volume, $n_2$ is the nonlinear refractive index, $c$ is the speed of light in a vacuum, $\eta$ is the coupling efficiency to the waveguide ($\eta=0.5$ is the critical coupling condition), and $g_R=6.2\times10^{-14}$ is the Raman gain coefficient of silica. For a silica microsphere with a diameter of 80~\textmu m, the parameters are assumed to be $n=1.44$, $n_2=2.2\times10^{-20}$, and $V_\mathrm{eff}=3287$~\textmu m$^3$. The effective mode volume is calculated by FEM simulation.

	\section*{Acknowledgments}
This work is supported by JSPS (KAKENHI JP22H01893,
JP22K14623, JP22K14624, JP22K14625, JP23H00262).
C.F.F. is supported by RIKEN Special
Postdoctoral Researcher Program. We thank the
Advanced Manufacturing Support Team at RIKEN for
technical assistance. We  also  thank  H.  Kumazaki  for preparing experimental setups.
\\
\\

	\section*{Author contributions}
S.F. and Y.K.K. conceived and designed the experiments. S.F. carried out sample preparation, numerical simulation, and experimental measurements. N.F. assisted the  materials transfer, and D.Y., D.K., and C.F.F. aided the construction of measurement setup. S.F. and Y.K.K. wrote the manuscript with inputs from all authors. Y.K.K. supervised the project.

	\bibliography{cavity_nanomaterial}

\begin{thebibliography}{48}%
\makeatletter
\providecommand \@ifxundefined [1]{%
 \@ifx{#1\undefined}
}%
\providecommand \@ifnum [1]{%
 \ifnum #1\expandafter \@firstoftwo
 \else \expandafter \@secondoftwo
 \fi
}%
\providecommand \@ifx [1]{%
 \ifx #1\expandafter \@firstoftwo
 \else \expandafter \@secondoftwo
 \fi
}%
\providecommand \natexlab [1]{#1}%
\providecommand \enquote  [1]{``#1''}%
\providecommand \bibnamefont  [1]{#1}%
\providecommand \bibfnamefont [1]{#1}%
\providecommand \citenamefont [1]{#1}%
\providecommand \href@noop [0]{\@secondoftwo}%
\providecommand \href [0]{\begingroup \@sanitize@url \@href}%
\providecommand \@href[1]{\@@startlink{#1}\@@href}%
\providecommand \@@href[1]{\endgroup#1\@@endlink}%
\providecommand \@sanitize@url [0]{\catcode `\\12\catcode `\$12\catcode
  `\&12\catcode `\#12\catcode `\^12\catcode `\_12\catcode `\%12\relax}%
\providecommand \@@startlink[1]{}%
\providecommand \@@endlink[0]{}%
\providecommand \url  [0]{\begingroup\@sanitize@url \@url }%
\providecommand \@url [1]{\endgroup\@href {#1}{\urlprefix }}%
\providecommand \urlprefix  [0]{URL }%
\providecommand \Eprint [0]{\href }%
\providecommand \doibase [0]{https://doi.org/}%
\providecommand \selectlanguage [0]{\@gobble}%
\providecommand \bibinfo  [0]{\@secondoftwo}%
\providecommand \bibfield  [0]{\@secondoftwo}%
\providecommand \translation [1]{[#1]}%
\providecommand \BibitemOpen [0]{}%
\providecommand \bibitemStop [0]{}%
\providecommand \bibitemNoStop [0]{.\EOS\space}%
\providecommand \EOS [0]{\spacefactor3000\relax}%
\providecommand \BibitemShut  [1]{\csname bibitem#1\endcsname}%
\let\auto@bib@innerbib\@empty
\bibitem [{\citenamefont {Franken}\ \emph {et~al.}(1961)\citenamefont
  {Franken}, \citenamefont {Hill}, \citenamefont {Peters},\ and\ \citenamefont
  {Weinreich}}]{Franken1961}%
  \BibitemOpen
  \bibfield  {author} {\bibinfo {author} {\bibfnamefont {P.~A.}\ \bibnamefont
  {Franken}}, \bibinfo {author} {\bibfnamefont {A.~E.}\ \bibnamefont {Hill}},
  \bibinfo {author} {\bibfnamefont {C.~W.}\ \bibnamefont {Peters}},\ and\
  \bibinfo {author} {\bibfnamefont {G.}~\bibnamefont {Weinreich}},\ }\bibfield
  {title} {\bibinfo {title} {Generation of optical harmonics},\ }\href
  {https://doi.org/10.1103/PhysRevLett.7.118} {\bibfield  {journal} {\bibinfo
  {journal} {Phys. Rev. Lett.}\ }\textbf {\bibinfo {volume} {7}},\ \bibinfo
  {pages} {118} (\bibinfo {year} {1961})}\BibitemShut {NoStop}%
\bibitem [{\citenamefont {Maiman}(1960)}]{MAIMAN1960}%
  \BibitemOpen
  \bibfield  {author} {\bibinfo {author} {\bibfnamefont {T.~H.}\ \bibnamefont
  {Maiman}},\ }\bibfield  {title} {\bibinfo {title} {Stimulated optical
  radiation in ruby},\ }\href {https://doi.org/10.1038/187493a0} {\bibfield
  {journal} {\bibinfo  {journal} {Nature}\ }\textbf {\bibinfo {volume} {187}},\
  \bibinfo {pages} {493} (\bibinfo {year} {1960})}\BibitemShut {NoStop}%
\bibitem [{\citenamefont {Armstrong}(2004)}]{Armstrong2004}%
  \BibitemOpen
  \bibfield  {author} {\bibinfo {author} {\bibfnamefont {J.~A.}\ \bibnamefont
  {Armstrong}},\ }\bibfield  {title} {\bibinfo {title} {Measurement of
  picosecond laser pulse widths},\ }\href {https://doi.org/10.1063/1.1754787}
  {\bibfield  {journal} {\bibinfo  {journal} {Appl. Phys. Lett.}\ }\textbf
  {\bibinfo {volume} {10}},\ \bibinfo {pages} {16} (\bibinfo {year}
  {2004})}\BibitemShut {NoStop}%
\bibitem [{\citenamefont {Reichert}\ \emph {et~al.}(1999)\citenamefont
  {Reichert}, \citenamefont {Holzwarth}, \citenamefont {Udem},\ and\
  \citenamefont {Hänsch}}]{Reichert1999}%
  \BibitemOpen
  \bibfield  {author} {\bibinfo {author} {\bibfnamefont {J.}~\bibnamefont
  {Reichert}}, \bibinfo {author} {\bibfnamefont {R.}~\bibnamefont {Holzwarth}},
  \bibinfo {author} {\bibfnamefont {T.}~\bibnamefont {Udem}},\ and\ \bibinfo
  {author} {\bibfnamefont {T.}~\bibnamefont {Hänsch}},\ }\bibfield  {title}
  {\bibinfo {title} {Measuring the frequency of light with mode-locked
  lasers},\ }\href
  {https://doi.org/https://doi.org/10.1016/S0030-4018(99)00491-5} {\bibfield
  {journal} {\bibinfo  {journal} {Opt. Commun.}\ }\textbf {\bibinfo {volume}
  {172}},\ \bibinfo {pages} {59} (\bibinfo {year} {1999})}\BibitemShut
  {NoStop}%
\bibitem [{\citenamefont {Jones}\ \emph {et~al.}(2000)\citenamefont {Jones},
  \citenamefont {Diddams}, \citenamefont {Ranka}, \citenamefont {Stentz},
  \citenamefont {Windeler}, \citenamefont {Hall},\ and\ \citenamefont
  {Cundiff}}]{Jones2000}%
  \BibitemOpen
  \bibfield  {author} {\bibinfo {author} {\bibfnamefont {D.~J.}\ \bibnamefont
  {Jones}}, \bibinfo {author} {\bibfnamefont {S.~A.}\ \bibnamefont {Diddams}},
  \bibinfo {author} {\bibfnamefont {J.~K.}\ \bibnamefont {Ranka}}, \bibinfo
  {author} {\bibfnamefont {A.}~\bibnamefont {Stentz}}, \bibinfo {author}
  {\bibfnamefont {R.~S.}\ \bibnamefont {Windeler}}, \bibinfo {author}
  {\bibfnamefont {J.~L.}\ \bibnamefont {Hall}},\ and\ \bibinfo {author}
  {\bibfnamefont {S.~T.}\ \bibnamefont {Cundiff}},\ }\bibfield  {title}
  {\bibinfo {title} {Carrier-envelope phase control of femtosecond mode-locked
  lasers and direct optical frequency synthesis},\ }\href
  {https://doi.org/10.1126/science.288.5466.635} {\bibfield  {journal}
  {\bibinfo  {journal} {Science}\ }\textbf {\bibinfo {volume} {288}},\ \bibinfo
  {pages} {635} (\bibinfo {year} {2000})}\BibitemShut {NoStop}%
\bibitem [{\citenamefont {Burnham}\ and\ \citenamefont
  {Weinberg}(1970)}]{Burnham1970}%
  \BibitemOpen
  \bibfield  {author} {\bibinfo {author} {\bibfnamefont {D.~C.}\ \bibnamefont
  {Burnham}}\ and\ \bibinfo {author} {\bibfnamefont {D.~L.}\ \bibnamefont
  {Weinberg}},\ }\bibfield  {title} {\bibinfo {title} {Observation of
  simultaneity in parametric production of optical photon pairs},\ }\href
  {https://doi.org/10.1103/PhysRevLett.25.84} {\bibfield  {journal} {\bibinfo
  {journal} {Phys. Rev. Lett.}\ }\textbf {\bibinfo {volume} {25}},\ \bibinfo
  {pages} {84} (\bibinfo {year} {1970})}\BibitemShut {NoStop}%
\bibitem [{\citenamefont {Slusher}\ \emph {et~al.}(1985)\citenamefont
  {Slusher}, \citenamefont {Hollberg}, \citenamefont {Yurke}, \citenamefont
  {Mertz},\ and\ \citenamefont {Valley}}]{Slusher1985}%
  \BibitemOpen
  \bibfield  {author} {\bibinfo {author} {\bibfnamefont {R.~E.}\ \bibnamefont
  {Slusher}}, \bibinfo {author} {\bibfnamefont {L.~W.}\ \bibnamefont
  {Hollberg}}, \bibinfo {author} {\bibfnamefont {B.}~\bibnamefont {Yurke}},
  \bibinfo {author} {\bibfnamefont {J.~C.}\ \bibnamefont {Mertz}},\ and\
  \bibinfo {author} {\bibfnamefont {J.~F.}\ \bibnamefont {Valley}},\ }\bibfield
   {title} {\bibinfo {title} {Observation of squeezed states generated by
  four-wave mixing in an optical cavity},\ }\href
  {https://doi.org/10.1103/PhysRevLett.55.2409} {\bibfield  {journal} {\bibinfo
   {journal} {Phys. Rev. Lett.}\ }\textbf {\bibinfo {volume} {55}},\ \bibinfo
  {pages} {2409} (\bibinfo {year} {1985})}\BibitemShut {NoStop}%
\bibitem [{\citenamefont {Squier}\ \emph {et~al.}(1998)\citenamefont {Squier},
  \citenamefont {M\"{u}ller}, \citenamefont {Brakenhoff},\ and\ \citenamefont
  {Wilson}}]{Squier1998}%
  \BibitemOpen
  \bibfield  {author} {\bibinfo {author} {\bibfnamefont {J.~A.}\ \bibnamefont
  {Squier}}, \bibinfo {author} {\bibfnamefont {M.}~\bibnamefont {M\"{u}ller}},
  \bibinfo {author} {\bibfnamefont {G.~J.}\ \bibnamefont {Brakenhoff}},\ and\
  \bibinfo {author} {\bibfnamefont {K.~R.}\ \bibnamefont {Wilson}},\ }\bibfield
   {title} {\bibinfo {title} {Third harmonic generation microscopy},\ }\href
  {https://doi.org/10.1364/OE.3.000315} {\bibfield  {journal} {\bibinfo
  {journal} {Opt. Express}\ }\textbf {\bibinfo {volume} {3}},\ \bibinfo {pages}
  {315} (\bibinfo {year} {1998})}\BibitemShut {NoStop}%
\bibitem [{\citenamefont {Shen}(1989)}]{Shen1989}%
  \BibitemOpen
  \bibfield  {author} {\bibinfo {author} {\bibfnamefont {Y.~R.}\ \bibnamefont
  {Shen}},\ }\bibfield  {title} {\bibinfo {title} {Surface properties probed by
  second-harmonic and sum-frequency generation},\ }\href
  {https://doi.org/10.1038/337519a0} {\bibfield  {journal} {\bibinfo  {journal}
  {Nature}\ }\textbf {\bibinfo {volume} {337}},\ \bibinfo {pages} {519}
  (\bibinfo {year} {1989})}\BibitemShut {NoStop}%
\bibitem [{\citenamefont {Boyd}(2003)}]{Boyd2003}%
  \BibitemOpen
  \bibfield  {author} {\bibinfo {author} {\bibfnamefont {R.~W.}\ \bibnamefont
  {Boyd}},\ }\href@noop {} {\emph {\bibinfo {title} {Nonlinear optics}}}\
  (\bibinfo  {publisher} {Taylor \& Francis},\ \bibinfo {year}
  {2003})\BibitemShut {NoStop}%
\bibitem [{\citenamefont {Fryett}\ \emph {et~al.}(2018)\citenamefont {Fryett},
  \citenamefont {Zhan},\ and\ \citenamefont {Majumdar}}]{Fryett2018}%
  \BibitemOpen
  \bibfield  {author} {\bibinfo {author} {\bibfnamefont {T.}~\bibnamefont
  {Fryett}}, \bibinfo {author} {\bibfnamefont {A.}~\bibnamefont {Zhan}},\ and\
  \bibinfo {author} {\bibfnamefont {A.}~\bibnamefont {Majumdar}},\ }\bibfield
  {title} {\bibinfo {title} {Cavity nonlinear optics with layered materials},\
  }\href {https://doi.org/doi:10.1515/nanoph-2017-0069} {\bibfield  {journal}
  {\bibinfo  {journal} {Nanophotonics}\ }\textbf {\bibinfo {volume} {7}},\
  \bibinfo {pages} {355} (\bibinfo {year} {2018})}\BibitemShut {NoStop}%
\bibitem [{\citenamefont {Kippenberg}\ \emph
  {et~al.}(2004{\natexlab{a}})\citenamefont {Kippenberg}, \citenamefont
  {Spillane},\ and\ \citenamefont {Vahala}}]{Kippenberg2004}%
  \BibitemOpen
  \bibfield  {author} {\bibinfo {author} {\bibfnamefont {T.~J.}\ \bibnamefont
  {Kippenberg}}, \bibinfo {author} {\bibfnamefont {S.~M.}\ \bibnamefont
  {Spillane}},\ and\ \bibinfo {author} {\bibfnamefont {K.~J.}\ \bibnamefont
  {Vahala}},\ }\bibfield  {title} {\bibinfo {title} {Kerr-nonlinearity optical
  parametric oscillation in an ultrahigh-{Q} toroid microcavity},\ }\href
  {https://doi.org/10.1103/PhysRevLett.93.083904} {\bibfield  {journal}
  {\bibinfo  {journal} {Phys. Rev. Lett.}\ }\textbf {\bibinfo {volume} {93}},\
  \bibinfo {pages} {083904} (\bibinfo {year} {2004}{\natexlab{a}})}\BibitemShut
  {NoStop}%
\bibitem [{\citenamefont {Kippenberg}\ \emph
  {et~al.}(2004{\natexlab{b}})\citenamefont {Kippenberg}, \citenamefont
  {Spillane}, \citenamefont {Min},\ and\ \citenamefont
  {Vahala}}]{Kippenberg2004a}%
  \BibitemOpen
  \bibfield  {author} {\bibinfo {author} {\bibfnamefont {T.~J.}\ \bibnamefont
  {Kippenberg}}, \bibinfo {author} {\bibfnamefont {S.~M.}\ \bibnamefont
  {Spillane}}, \bibinfo {author} {\bibfnamefont {B.}~\bibnamefont {Min}},\ and\
  \bibinfo {author} {\bibfnamefont {K.~J.}\ \bibnamefont {Vahala}},\ }\bibfield
   {title} {\bibinfo {title} {Theoretical and experimental study of stimulated
  and cascaded {R}aman scattering in ultrahigh-{Q} optical microcavities},\
  }\href {https://doi.org/10.1109/JSTQE.2004.837203} {\bibfield  {journal}
  {\bibinfo  {journal} {IEEE J. Sel. Topics Quantum Electron.}\ }\textbf
  {\bibinfo {volume} {10}},\ \bibinfo {pages} {1219} (\bibinfo {year}
  {2004}{\natexlab{b}})}\BibitemShut {NoStop}%
\bibitem [{\citenamefont {Del'Haye}\ \emph {et~al.}(2013)\citenamefont
  {Del'Haye}, \citenamefont {Diddams},\ and\ \citenamefont
  {Papp}}]{DelHaye2013}%
  \BibitemOpen
  \bibfield  {author} {\bibinfo {author} {\bibfnamefont {P.}~\bibnamefont
  {Del'Haye}}, \bibinfo {author} {\bibfnamefont {S.~A.}\ \bibnamefont
  {Diddams}},\ and\ \bibinfo {author} {\bibfnamefont {S.~B.}\ \bibnamefont
  {Papp}},\ }\bibfield  {title} {\bibinfo {title} {Laser-machined
  ultra-high-{Q} microrod resonators for nonlinear optics},\ }\href
  {https://doi.org/10.1063/1.4809781} {\bibfield  {journal} {\bibinfo
  {journal} {Appl. Phys. Lett.}\ }\textbf {\bibinfo {volume} {102}},\ \bibinfo
  {pages} {221119} (\bibinfo {year} {2013})}\BibitemShut {NoStop}%
\bibitem [{\citenamefont {Farnesi}\ \emph {et~al.}(2014)\citenamefont
  {Farnesi}, \citenamefont {Barucci}, \citenamefont {Righini}, \citenamefont
  {Berneschi}, \citenamefont {Soria},\ and\ \citenamefont
  {Nunzi~Conti}}]{Farnesi2014}%
  \BibitemOpen
  \bibfield  {author} {\bibinfo {author} {\bibfnamefont {D.}~\bibnamefont
  {Farnesi}}, \bibinfo {author} {\bibfnamefont {A.}~\bibnamefont {Barucci}},
  \bibinfo {author} {\bibfnamefont {G.~C.}\ \bibnamefont {Righini}}, \bibinfo
  {author} {\bibfnamefont {S.}~\bibnamefont {Berneschi}}, \bibinfo {author}
  {\bibfnamefont {S.}~\bibnamefont {Soria}},\ and\ \bibinfo {author}
  {\bibfnamefont {G.}~\bibnamefont {Nunzi~Conti}},\ }\bibfield  {title}
  {\bibinfo {title} {Optical frequency conversion in
  silica-whispering-gallery-mode microspherical resonators},\ }\href
  {https://doi.org/10.1103/PhysRevLett.112.093901} {\bibfield  {journal}
  {\bibinfo  {journal} {Phys. Rev. Lett.}\ }\textbf {\bibinfo {volume} {112}},\
  \bibinfo {pages} {093901} (\bibinfo {year} {2014})}\BibitemShut {NoStop}%
\bibitem [{\citenamefont {Yi}\ \emph {et~al.}(2015)\citenamefont {Yi},
  \citenamefont {Yang}, \citenamefont {Yang}, \citenamefont {Suh},\ and\
  \citenamefont {Vahala}}]{Yi2015}%
  \BibitemOpen
  \bibfield  {author} {\bibinfo {author} {\bibfnamefont {X.}~\bibnamefont
  {Yi}}, \bibinfo {author} {\bibfnamefont {Q.-F.}\ \bibnamefont {Yang}},
  \bibinfo {author} {\bibfnamefont {K.~Y.}\ \bibnamefont {Yang}}, \bibinfo
  {author} {\bibfnamefont {M.-G.}\ \bibnamefont {Suh}},\ and\ \bibinfo {author}
  {\bibfnamefont {K.}~\bibnamefont {Vahala}},\ }\bibfield  {title} {\bibinfo
  {title} {Soliton frequency comb at microwave rates in a high-{Q} silica
  microresonator},\ }\href {https://doi.org/10.1364/OPTICA.2.001078} {\bibfield
   {journal} {\bibinfo  {journal} {Optica}\ }\textbf {\bibinfo {volume} {2}},\
  \bibinfo {pages} {1078} (\bibinfo {year} {2015})}\BibitemShut {NoStop}%
\bibitem [{\citenamefont {Chen-Jinnai}\ \emph {et~al.}(2016)\citenamefont
  {Chen-Jinnai}, \citenamefont {Kato}, \citenamefont {Fujii}, \citenamefont
  {Nagano}, \citenamefont {Kobatake},\ and\ \citenamefont
  {Tanabe}}]{ChenJinnai2016}%
  \BibitemOpen
  \bibfield  {author} {\bibinfo {author} {\bibfnamefont {A.}~\bibnamefont
  {Chen-Jinnai}}, \bibinfo {author} {\bibfnamefont {T.}~\bibnamefont {Kato}},
  \bibinfo {author} {\bibfnamefont {S.}~\bibnamefont {Fujii}}, \bibinfo
  {author} {\bibfnamefont {T.}~\bibnamefont {Nagano}}, \bibinfo {author}
  {\bibfnamefont {T.}~\bibnamefont {Kobatake}},\ and\ \bibinfo {author}
  {\bibfnamefont {T.}~\bibnamefont {Tanabe}},\ }\bibfield  {title} {\bibinfo
  {title} {Broad bandwidth third-harmonic generation via four-wave mixing and
  stimulated {R}aman scattering in a microcavity},\ }\href@noop {} {\bibfield
  {journal} {\bibinfo  {journal} {Opt. Express}\ }\textbf {\bibinfo {volume}
  {24}},\ \bibinfo {pages} {26322} (\bibinfo {year} {2016})}\BibitemShut
  {NoStop}%
\bibitem [{\citenamefont {Fujii}\ \emph {et~al.}(2017)\citenamefont {Fujii},
  \citenamefont {Kato}, \citenamefont {Suzuki},\ and\ \citenamefont
  {Tanabe}}]{Fujii2017}%
  \BibitemOpen
  \bibfield  {author} {\bibinfo {author} {\bibfnamefont {S.}~\bibnamefont
  {Fujii}}, \bibinfo {author} {\bibfnamefont {T.}~\bibnamefont {Kato}},
  \bibinfo {author} {\bibfnamefont {R.}~\bibnamefont {Suzuki}},\ and\ \bibinfo
  {author} {\bibfnamefont {T.}~\bibnamefont {Tanabe}},\ }\bibfield  {title}
  {\bibinfo {title} {Third-harmonic blue light generation from {K}err clustered
  combs and dispersive waves},\ }\href {https://doi.org/10.1364/OL.42.002010}
  {\bibfield  {journal} {\bibinfo  {journal} {Opt. Lett.}\ }\textbf {\bibinfo
  {volume} {42}},\ \bibinfo {pages} {2010} (\bibinfo {year}
  {2017})}\BibitemShut {NoStop}%
\bibitem [{\citenamefont {Malard}\ \emph {et~al.}(2013)\citenamefont {Malard},
  \citenamefont {Alencar}, \citenamefont {Barboza}, \citenamefont {Mak},\ and\
  \citenamefont {de~Paula}}]{Malard2013}%
  \BibitemOpen
  \bibfield  {author} {\bibinfo {author} {\bibfnamefont {L.~M.}\ \bibnamefont
  {Malard}}, \bibinfo {author} {\bibfnamefont {T.~V.}\ \bibnamefont {Alencar}},
  \bibinfo {author} {\bibfnamefont {A.~P.~M.}\ \bibnamefont {Barboza}},
  \bibinfo {author} {\bibfnamefont {K.~F.}\ \bibnamefont {Mak}},\ and\ \bibinfo
  {author} {\bibfnamefont {A.~M.}\ \bibnamefont {de~Paula}},\ }\bibfield
  {title} {\bibinfo {title} {Observation of intense second harmonic generation
  from {M}o{S}$_{2}$ atomic crystals},\ }\href
  {https://doi.org/10.1103/PhysRevB.87.201401} {\bibfield  {journal} {\bibinfo
  {journal} {Phys. Rev. B}\ }\textbf {\bibinfo {volume} {87}},\ \bibinfo
  {pages} {201401} (\bibinfo {year} {2013})}\BibitemShut {NoStop}%
\bibitem [{\citenamefont {Woodward}\ \emph {et~al.}(2016)\citenamefont
  {Woodward}, \citenamefont {Murray}, \citenamefont {Phelan}, \citenamefont
  {de~Oliveira}, \citenamefont {Runcorn}, \citenamefont {Kelleher},
  \citenamefont {Li}, \citenamefont {de~Oliveira}, \citenamefont {Fechine},
  \citenamefont {Eda},\ and\ \citenamefont {de~Matos}}]{Woodward2016}%
  \BibitemOpen
  \bibfield  {author} {\bibinfo {author} {\bibfnamefont {R.~I.}\ \bibnamefont
  {Woodward}}, \bibinfo {author} {\bibfnamefont {R.~T.}\ \bibnamefont
  {Murray}}, \bibinfo {author} {\bibfnamefont {C.~F.}\ \bibnamefont {Phelan}},
  \bibinfo {author} {\bibfnamefont {R.~E.~P.}\ \bibnamefont {de~Oliveira}},
  \bibinfo {author} {\bibfnamefont {T.~H.}\ \bibnamefont {Runcorn}}, \bibinfo
  {author} {\bibfnamefont {E.~J.~R.}\ \bibnamefont {Kelleher}}, \bibinfo
  {author} {\bibfnamefont {S.}~\bibnamefont {Li}}, \bibinfo {author}
  {\bibfnamefont {E.~C.}\ \bibnamefont {de~Oliveira}}, \bibinfo {author}
  {\bibfnamefont {G.~J.~M.}\ \bibnamefont {Fechine}}, \bibinfo {author}
  {\bibfnamefont {G.}~\bibnamefont {Eda}},\ and\ \bibinfo {author}
  {\bibfnamefont {C.~J.~S.}\ \bibnamefont {de~Matos}},\ }\bibfield  {title}
  {\bibinfo {title} {Characterization of the second- and third-order nonlinear
  optical susceptibilities of monolayer {M}o{S}$_2$ using multiphoton
  microscopy},\ }\href {https://doi.org/10.1088/2053-1583/4/1/011006}
  {\bibfield  {journal} {\bibinfo  {journal} {2D Mater.}\ }\textbf {\bibinfo
  {volume} {4}},\ \bibinfo {pages} {011006} (\bibinfo {year}
  {2016})}\BibitemShut {NoStop}%
\bibitem [{\citenamefont {Autere}\ \emph {et~al.}(2018)\citenamefont {Autere},
  \citenamefont {Jussila}, \citenamefont {Marini}, \citenamefont {Saavedra},
  \citenamefont {Dai}, \citenamefont {S\"ayn\"atjoki}, \citenamefont
  {Karvonen}, \citenamefont {Yang}, \citenamefont {Amirsolaimani},
  \citenamefont {Norwood}, \citenamefont {Peyghambarian}, \citenamefont
  {Lipsanen}, \citenamefont {Kieu}, \citenamefont {de~Abajo},\ and\
  \citenamefont {Sun}}]{Autere2018}%
  \BibitemOpen
  \bibfield  {author} {\bibinfo {author} {\bibfnamefont {A.}~\bibnamefont
  {Autere}}, \bibinfo {author} {\bibfnamefont {H.}~\bibnamefont {Jussila}},
  \bibinfo {author} {\bibfnamefont {A.}~\bibnamefont {Marini}}, \bibinfo
  {author} {\bibfnamefont {J.~R.~M.}\ \bibnamefont {Saavedra}}, \bibinfo
  {author} {\bibfnamefont {Y.}~\bibnamefont {Dai}}, \bibinfo {author}
  {\bibfnamefont {A.}~\bibnamefont {S\"ayn\"atjoki}}, \bibinfo {author}
  {\bibfnamefont {L.}~\bibnamefont {Karvonen}}, \bibinfo {author}
  {\bibfnamefont {H.}~\bibnamefont {Yang}}, \bibinfo {author} {\bibfnamefont
  {B.}~\bibnamefont {Amirsolaimani}}, \bibinfo {author} {\bibfnamefont {R.~A.}\
  \bibnamefont {Norwood}}, \bibinfo {author} {\bibfnamefont {N.}~\bibnamefont
  {Peyghambarian}}, \bibinfo {author} {\bibfnamefont {H.}~\bibnamefont
  {Lipsanen}}, \bibinfo {author} {\bibfnamefont {K.}~\bibnamefont {Kieu}},
  \bibinfo {author} {\bibfnamefont {F.~J.~G.}\ \bibnamefont {de~Abajo}},\ and\
  \bibinfo {author} {\bibfnamefont {Z.}~\bibnamefont {Sun}},\ }\bibfield
  {title} {\bibinfo {title} {Optical harmonic generation in monolayer
  group-{VI} transition metal dichalcogenides},\ }\href
  {https://doi.org/10.1103/PhysRevB.98.115426} {\bibfield  {journal} {\bibinfo
  {journal} {Phys. Rev. B}\ }\textbf {\bibinfo {volume} {98}},\ \bibinfo
  {pages} {115426} (\bibinfo {year} {2018})}\BibitemShut {NoStop}%
\bibitem [{\citenamefont {Lin}\ \emph {et~al.}(2019)\citenamefont {Lin},
  \citenamefont {Bange},\ and\ \citenamefont {Lupton}}]{Lin2019}%
  \BibitemOpen
  \bibfield  {author} {\bibinfo {author} {\bibfnamefont {K.-Q.}\ \bibnamefont
  {Lin}}, \bibinfo {author} {\bibfnamefont {S.}~\bibnamefont {Bange}},\ and\
  \bibinfo {author} {\bibfnamefont {J.~M.}\ \bibnamefont {Lupton}},\ }\bibfield
   {title} {\bibinfo {title} {Quantum interference in second-harmonic
  generation from monolayer {WS}e$_2$},\ }\href
  {https://doi.org/10.1038/s41567-018-0384-5} {\bibfield  {journal} {\bibinfo
  {journal} {Nat. Phys.}\ }\textbf {\bibinfo {volume} {15}},\ \bibinfo {pages}
  {242} (\bibinfo {year} {2019})}\BibitemShut {NoStop}%
\bibitem [{\citenamefont {Klimmer}\ \emph {et~al.}(2021)\citenamefont
  {Klimmer}, \citenamefont {Ghaebi}, \citenamefont {Gan}, \citenamefont
  {George}, \citenamefont {Turchanin}, \citenamefont {Cerullo},\ and\
  \citenamefont {Soavi}}]{Klimmer2021}%
  \BibitemOpen
  \bibfield  {author} {\bibinfo {author} {\bibfnamefont {S.}~\bibnamefont
  {Klimmer}}, \bibinfo {author} {\bibfnamefont {O.}~\bibnamefont {Ghaebi}},
  \bibinfo {author} {\bibfnamefont {Z.}~\bibnamefont {Gan}}, \bibinfo {author}
  {\bibfnamefont {A.}~\bibnamefont {George}}, \bibinfo {author} {\bibfnamefont
  {A.}~\bibnamefont {Turchanin}}, \bibinfo {author} {\bibfnamefont
  {G.}~\bibnamefont {Cerullo}},\ and\ \bibinfo {author} {\bibfnamefont
  {G.}~\bibnamefont {Soavi}},\ }\bibfield  {title} {\bibinfo {title}
  {All-optical polarization and amplitude modulation of second-harmonic
  generation in atomically thin semiconductors},\ }\href
  {https://doi.org/10.1038/s41566-021-00859-y} {\bibfield  {journal} {\bibinfo
  {journal} {Nat. Photon.}\ }\textbf {\bibinfo {volume} {15}},\ \bibinfo
  {pages} {837} (\bibinfo {year} {2021})}\BibitemShut {NoStop}%
\bibitem [{\citenamefont {Trovatello}\ \emph {et~al.}(2021)\citenamefont
  {Trovatello}, \citenamefont {Marini}, \citenamefont {Xu}, \citenamefont
  {Lee}, \citenamefont {Liu}, \citenamefont {Curreli}, \citenamefont {Manzoni},
  \citenamefont {Dal~Conte}, \citenamefont {Yao}, \citenamefont {Ciattoni},
  \citenamefont {Hone}, \citenamefont {Zhu}, \citenamefont {Schuck},\ and\
  \citenamefont {Cerullo}}]{Trovatello2021}%
  \BibitemOpen
  \bibfield  {author} {\bibinfo {author} {\bibfnamefont {C.}~\bibnamefont
  {Trovatello}}, \bibinfo {author} {\bibfnamefont {A.}~\bibnamefont {Marini}},
  \bibinfo {author} {\bibfnamefont {X.}~\bibnamefont {Xu}}, \bibinfo {author}
  {\bibfnamefont {C.}~\bibnamefont {Lee}}, \bibinfo {author} {\bibfnamefont
  {F.}~\bibnamefont {Liu}}, \bibinfo {author} {\bibfnamefont {N.}~\bibnamefont
  {Curreli}}, \bibinfo {author} {\bibfnamefont {C.}~\bibnamefont {Manzoni}},
  \bibinfo {author} {\bibfnamefont {S.}~\bibnamefont {Dal~Conte}}, \bibinfo
  {author} {\bibfnamefont {K.}~\bibnamefont {Yao}}, \bibinfo {author}
  {\bibfnamefont {A.}~\bibnamefont {Ciattoni}}, \bibinfo {author}
  {\bibfnamefont {J.}~\bibnamefont {Hone}}, \bibinfo {author} {\bibfnamefont
  {X.}~\bibnamefont {Zhu}}, \bibinfo {author} {\bibfnamefont {P.~J.}\
  \bibnamefont {Schuck}},\ and\ \bibinfo {author} {\bibfnamefont
  {G.}~\bibnamefont {Cerullo}},\ }\bibfield  {title} {\bibinfo {title} {Optical
  parametric amplification by monolayer transition metal dichalcogenides},\
  }\href {https://doi.org/10.1038/s41566-020-00728-0} {\bibfield  {journal}
  {\bibinfo  {journal} {Nat. Photon.}\ }\textbf {\bibinfo {volume} {15}},\
  \bibinfo {pages} {6} (\bibinfo {year} {2021})}\BibitemShut {NoStop}%
\bibitem [{\citenamefont {Xia}\ \emph {et~al.}(2014)\citenamefont {Xia},
  \citenamefont {Wang}, \citenamefont {Xiao}, \citenamefont {Dubey},\ and\
  \citenamefont {Ramasubramaniam}}]{Xia2014}%
  \BibitemOpen
  \bibfield  {author} {\bibinfo {author} {\bibfnamefont {F.}~\bibnamefont
  {Xia}}, \bibinfo {author} {\bibfnamefont {H.}~\bibnamefont {Wang}}, \bibinfo
  {author} {\bibfnamefont {D.}~\bibnamefont {Xiao}}, \bibinfo {author}
  {\bibfnamefont {M.}~\bibnamefont {Dubey}},\ and\ \bibinfo {author}
  {\bibfnamefont {A.}~\bibnamefont {Ramasubramaniam}},\ }\bibfield  {title}
  {\bibinfo {title} {Two-dimensional material nanophotonics},\ }\href
  {https://doi.org/10.1038/nphoton.2014.271} {\bibfield  {journal} {\bibinfo
  {journal} {Nat. Photon.}\ }\textbf {\bibinfo {volume} {8}},\ \bibinfo {pages}
  {899} (\bibinfo {year} {2014})}\BibitemShut {NoStop}%
\bibitem [{\citenamefont {Ngo}\ \emph {et~al.}(2022)\citenamefont {Ngo},
  \citenamefont {Najafidehaghani}, \citenamefont {Gan}, \citenamefont
  {Khazaee}, \citenamefont {Siems}, \citenamefont {George}, \citenamefont
  {Schartner}, \citenamefont {Nolte}, \citenamefont {Ebendorff-Heidepriem},
  \citenamefont {Pertsch}, \citenamefont {Tuniz}, \citenamefont {Schmidt},
  \citenamefont {Peschel}, \citenamefont {Turchanin},\ and\ \citenamefont
  {Eilenberger}}]{Ngo2022}%
  \BibitemOpen
  \bibfield  {author} {\bibinfo {author} {\bibfnamefont {G.~Q.}\ \bibnamefont
  {Ngo}}, \bibinfo {author} {\bibfnamefont {E.}~\bibnamefont
  {Najafidehaghani}}, \bibinfo {author} {\bibfnamefont {Z.}~\bibnamefont
  {Gan}}, \bibinfo {author} {\bibfnamefont {S.}~\bibnamefont {Khazaee}},
  \bibinfo {author} {\bibfnamefont {M.~P.}\ \bibnamefont {Siems}}, \bibinfo
  {author} {\bibfnamefont {A.}~\bibnamefont {George}}, \bibinfo {author}
  {\bibfnamefont {E.~P.}\ \bibnamefont {Schartner}}, \bibinfo {author}
  {\bibfnamefont {S.}~\bibnamefont {Nolte}}, \bibinfo {author} {\bibfnamefont
  {H.}~\bibnamefont {Ebendorff-Heidepriem}}, \bibinfo {author} {\bibfnamefont
  {T.}~\bibnamefont {Pertsch}}, \bibinfo {author} {\bibfnamefont
  {A.}~\bibnamefont {Tuniz}}, \bibinfo {author} {\bibfnamefont {M.~A.}\
  \bibnamefont {Schmidt}}, \bibinfo {author} {\bibfnamefont {U.}~\bibnamefont
  {Peschel}}, \bibinfo {author} {\bibfnamefont {A.}~\bibnamefont {Turchanin}},\
  and\ \bibinfo {author} {\bibfnamefont {F.}~\bibnamefont {Eilenberger}},\
  }\bibfield  {title} {\bibinfo {title} {In-fibre second-harmonic generation
  with embedded two-dimensional materials},\ }\href
  {https://doi.org/10.1038/s41566-022-01067-y} {\bibfield  {journal} {\bibinfo
  {journal} {Nat. Photon.}\ }\textbf {\bibinfo {volume} {16}},\ \bibinfo
  {pages} {769} (\bibinfo {year} {2022})}\BibitemShut {NoStop}%
\bibitem [{\citenamefont {Javerzac-Galy}\ \emph {et~al.}(2018)\citenamefont
  {Javerzac-Galy}, \citenamefont {Kumar}, \citenamefont {Schilling},
  \citenamefont {Piro}, \citenamefont {Khorasani}, \citenamefont {Barbone},
  \citenamefont {Goykhman}, \citenamefont {Khurgin}, \citenamefont {Ferrari},\
  and\ \citenamefont {Kippenberg}}]{Javerzac-Galy2018}%
  \BibitemOpen
  \bibfield  {author} {\bibinfo {author} {\bibfnamefont {C.}~\bibnamefont
  {Javerzac-Galy}}, \bibinfo {author} {\bibfnamefont {A.}~\bibnamefont
  {Kumar}}, \bibinfo {author} {\bibfnamefont {R.~D.}\ \bibnamefont
  {Schilling}}, \bibinfo {author} {\bibfnamefont {N.}~\bibnamefont {Piro}},
  \bibinfo {author} {\bibfnamefont {S.}~\bibnamefont {Khorasani}}, \bibinfo
  {author} {\bibfnamefont {M.}~\bibnamefont {Barbone}}, \bibinfo {author}
  {\bibfnamefont {I.}~\bibnamefont {Goykhman}}, \bibinfo {author}
  {\bibfnamefont {J.~B.}\ \bibnamefont {Khurgin}}, \bibinfo {author}
  {\bibfnamefont {A.~C.}\ \bibnamefont {Ferrari}},\ and\ \bibinfo {author}
  {\bibfnamefont {T.~J.}\ \bibnamefont {Kippenberg}},\ }\bibfield  {title}
  {\bibinfo {title} {Excitonic emission of monolayer semiconductors near-field
  coupled to high-{Q} microresonators},\ }\href
  {https://doi.org/10.1021/acs.nanolett.8b00749} {\bibfield  {journal}
  {\bibinfo  {journal} {Nano Lett.}\ }\textbf {\bibinfo {volume} {18}},\
  \bibinfo {pages} {3138} (\bibinfo {year} {2018})}\BibitemShut {NoStop}%
\bibitem [{\citenamefont {Tan}\ \emph {et~al.}(2021)\citenamefont {Tan},
  \citenamefont {Yuan}, \citenamefont {Zhang}, \citenamefont {Yan},
  \citenamefont {Zhou}, \citenamefont {An}, \citenamefont {Peng}, \citenamefont
  {Soavi}, \citenamefont {Rao},\ and\ \citenamefont {Yao}}]{Tan2021}%
  \BibitemOpen
  \bibfield  {author} {\bibinfo {author} {\bibfnamefont {T.}~\bibnamefont
  {Tan}}, \bibinfo {author} {\bibfnamefont {Z.}~\bibnamefont {Yuan}}, \bibinfo
  {author} {\bibfnamefont {H.}~\bibnamefont {Zhang}}, \bibinfo {author}
  {\bibfnamefont {G.}~\bibnamefont {Yan}}, \bibinfo {author} {\bibfnamefont
  {S.}~\bibnamefont {Zhou}}, \bibinfo {author} {\bibfnamefont {N.}~\bibnamefont
  {An}}, \bibinfo {author} {\bibfnamefont {B.}~\bibnamefont {Peng}}, \bibinfo
  {author} {\bibfnamefont {G.}~\bibnamefont {Soavi}}, \bibinfo {author}
  {\bibfnamefont {Y.}~\bibnamefont {Rao}},\ and\ \bibinfo {author}
  {\bibfnamefont {B.}~\bibnamefont {Yao}},\ }\bibfield  {title} {\bibinfo
  {title} {Multispecies and individual gas molecule detection using {S}tokes
  solitons in a graphene over-modal microresonator},\ }\href
  {https://doi.org/10.1038/s41467-021-26740-8} {\bibfield  {journal} {\bibinfo
  {journal} {Nat. Commun.}\ }\textbf {\bibinfo {volume} {12}},\ \bibinfo
  {pages} {6716} (\bibinfo {year} {2021})}\BibitemShut {NoStop}%
\bibitem [{\citenamefont {He}\ \emph {et~al.}(2021)\citenamefont {He},
  \citenamefont {Paradisanos}, \citenamefont {Liu}, \citenamefont {Cadore},
  \citenamefont {Liu}, \citenamefont {Churaev}, \citenamefont {Wang},
  \citenamefont {Raja}, \citenamefont {Javerzac-Galy}, \citenamefont {Roelli},
  \citenamefont {Fazio}, \citenamefont {Rosa}, \citenamefont {Tongay},
  \citenamefont {Soavi}, \citenamefont {Ferrari},\ and\ \citenamefont
  {Kippenberg}}]{He2021}%
  \BibitemOpen
  \bibfield  {author} {\bibinfo {author} {\bibfnamefont {J.}~\bibnamefont
  {He}}, \bibinfo {author} {\bibfnamefont {I.}~\bibnamefont {Paradisanos}},
  \bibinfo {author} {\bibfnamefont {T.}~\bibnamefont {Liu}}, \bibinfo {author}
  {\bibfnamefont {A.~R.}\ \bibnamefont {Cadore}}, \bibinfo {author}
  {\bibfnamefont {J.}~\bibnamefont {Liu}}, \bibinfo {author} {\bibfnamefont
  {M.}~\bibnamefont {Churaev}}, \bibinfo {author} {\bibfnamefont {R.~N.}\
  \bibnamefont {Wang}}, \bibinfo {author} {\bibfnamefont {A.~S.}\ \bibnamefont
  {Raja}}, \bibinfo {author} {\bibfnamefont {C.}~\bibnamefont {Javerzac-Galy}},
  \bibinfo {author} {\bibfnamefont {P.}~\bibnamefont {Roelli}}, \bibinfo
  {author} {\bibfnamefont {D.~D.}\ \bibnamefont {Fazio}}, \bibinfo {author}
  {\bibfnamefont {B.~L.~T.}\ \bibnamefont {Rosa}}, \bibinfo {author}
  {\bibfnamefont {S.}~\bibnamefont {Tongay}}, \bibinfo {author} {\bibfnamefont
  {G.}~\bibnamefont {Soavi}}, \bibinfo {author} {\bibfnamefont {A.~C.}\
  \bibnamefont {Ferrari}},\ and\ \bibinfo {author} {\bibfnamefont {T.~J.}\
  \bibnamefont {Kippenberg}},\ }\bibfield  {title} {\bibinfo {title} {Low-loss
  integrated nanophotonic circuits with layered semiconductor materials},\
  }\href {https://doi.org/10.1021/acs.nanolett.0c04149} {\bibfield  {journal}
  {\bibinfo  {journal} {Nano Lett.}\ }\textbf {\bibinfo {volume} {21}},\
  \bibinfo {pages} {2709} (\bibinfo {year} {2021})}\BibitemShut {NoStop}%
\bibitem [{\citenamefont {Fang}\ \emph {et~al.}(2022)\citenamefont {Fang},
  \citenamefont {Yamashita}, \citenamefont {Fujii}, \citenamefont {Otsuka},
  \citenamefont {Taniguchi}, \citenamefont {Watanabe}, \citenamefont
  {Nagashio},\ and\ \citenamefont {Kato}}]{Nan2023}%
  \BibitemOpen
  \bibfield  {author} {\bibinfo {author} {\bibfnamefont {N.}~\bibnamefont
  {Fang}}, \bibinfo {author} {\bibfnamefont {D.}~\bibnamefont {Yamashita}},
  \bibinfo {author} {\bibfnamefont {S.}~\bibnamefont {Fujii}}, \bibinfo
  {author} {\bibfnamefont {K.}~\bibnamefont {Otsuka}}, \bibinfo {author}
  {\bibfnamefont {T.}~\bibnamefont {Taniguchi}}, \bibinfo {author}
  {\bibfnamefont {K.}~\bibnamefont {Watanabe}}, \bibinfo {author}
  {\bibfnamefont {K.}~\bibnamefont {Nagashio}},\ and\ \bibinfo {author}
  {\bibfnamefont {Y.~K.}\ \bibnamefont {Kato}},\ }\bibfield  {title} {\bibinfo
  {title} {Quantization of mode shifts in nanocavities integrated with
  atomically thin sheets},\ }\href
  {https://doi.org/https://doi.org/10.1002/adom.202200538} {\bibfield
  {journal} {\bibinfo  {journal} {Adv. Opt. Mater.}\ }\textbf {\bibinfo
  {volume} {10}},\ \bibinfo {pages} {2200538} (\bibinfo {year}
  {2022})}\BibitemShut {NoStop}%
\bibitem [{\citenamefont {Castellanos-Gomez}\ \emph {et~al.}(2014)\citenamefont
  {Castellanos-Gomez}, \citenamefont {Buscema}, \citenamefont {Molenaar},
  \citenamefont {Singh}, \citenamefont {Janssen}, \citenamefont {van~der
  Zant},\ and\ \citenamefont {Steele}}]{CastellanosGomez2014}%
  \BibitemOpen
  \bibfield  {author} {\bibinfo {author} {\bibfnamefont {A.}~\bibnamefont
  {Castellanos-Gomez}}, \bibinfo {author} {\bibfnamefont {M.}~\bibnamefont
  {Buscema}}, \bibinfo {author} {\bibfnamefont {R.}~\bibnamefont {Molenaar}},
  \bibinfo {author} {\bibfnamefont {V.}~\bibnamefont {Singh}}, \bibinfo
  {author} {\bibfnamefont {L.}~\bibnamefont {Janssen}}, \bibinfo {author}
  {\bibfnamefont {H.~S.~J.}\ \bibnamefont {van~der Zant}},\ and\ \bibinfo
  {author} {\bibfnamefont {G.~A.}\ \bibnamefont {Steele}},\ }\bibfield  {title}
  {\bibinfo {title} {Deterministic transfer of two-dimensional materials by
  all-dry viscoelastic stamping},\ }\href
  {https://doi.org/10.1088/2053-1583/1/1/011002} {\bibfield  {journal}
  {\bibinfo  {journal} {2D Mater.}\ }\textbf {\bibinfo {volume} {1}},\ \bibinfo
  {pages} {011002} (\bibinfo {year} {2014})}\BibitemShut {NoStop}%
\bibitem [{\citenamefont {Zhao}\ \emph {et~al.}(2013)\citenamefont {Zhao},
  \citenamefont {Ghorannevis}, \citenamefont {Chu}, \citenamefont {Toh},
  \citenamefont {Kloc}, \citenamefont {Tan},\ and\ \citenamefont
  {Eda}}]{Zhao2013}%
  \BibitemOpen
  \bibfield  {author} {\bibinfo {author} {\bibfnamefont {W.}~\bibnamefont
  {Zhao}}, \bibinfo {author} {\bibfnamefont {Z.}~\bibnamefont {Ghorannevis}},
  \bibinfo {author} {\bibfnamefont {L.}~\bibnamefont {Chu}}, \bibinfo {author}
  {\bibfnamefont {M.}~\bibnamefont {Toh}}, \bibinfo {author} {\bibfnamefont
  {C.}~\bibnamefont {Kloc}}, \bibinfo {author} {\bibfnamefont {P.-H.}\
  \bibnamefont {Tan}},\ and\ \bibinfo {author} {\bibfnamefont {G.}~\bibnamefont
  {Eda}},\ }\bibfield  {title} {\bibinfo {title} {Evolution of electronic
  structure in atomically thin sheets of {WS}$_2$ and {WS}e$_2$},\ }\href
  {https://doi.org/10.1021/nn305275h} {\bibfield  {journal} {\bibinfo
  {journal} {ACS Nano}\ }\textbf {\bibinfo {volume} {7}},\ \bibinfo {pages}
  {791} (\bibinfo {year} {2013})}\BibitemShut {NoStop}%
\bibitem [{\citenamefont {Li}\ \emph {et~al.}(2013{\natexlab{a}})\citenamefont
  {Li}, \citenamefont {Wu}, \citenamefont {Huang}, \citenamefont {Lu},
  \citenamefont {Yang}, \citenamefont {Lu}, \citenamefont {Xiong},\ and\
  \citenamefont {Zhang}}]{Li2013rapid}%
  \BibitemOpen
  \bibfield  {author} {\bibinfo {author} {\bibfnamefont {H.}~\bibnamefont
  {Li}}, \bibinfo {author} {\bibfnamefont {J.}~\bibnamefont {Wu}}, \bibinfo
  {author} {\bibfnamefont {X.}~\bibnamefont {Huang}}, \bibinfo {author}
  {\bibfnamefont {G.}~\bibnamefont {Lu}}, \bibinfo {author} {\bibfnamefont
  {J.}~\bibnamefont {Yang}}, \bibinfo {author} {\bibfnamefont {X.}~\bibnamefont
  {Lu}}, \bibinfo {author} {\bibfnamefont {Q.}~\bibnamefont {Xiong}},\ and\
  \bibinfo {author} {\bibfnamefont {H.}~\bibnamefont {Zhang}},\ }\bibfield
  {title} {\bibinfo {title} {Rapid and reliable thickness identification of
  two-dimensional nanosheets using optical microscopy},\ }\href
  {https://doi.org/10.1021/nn4047474} {\bibfield  {journal} {\bibinfo
  {journal} {ACS Nano}\ }\textbf {\bibinfo {volume} {7}},\ \bibinfo {pages}
  {10344} (\bibinfo {year} {2013}{\natexlab{a}})}\BibitemShut {NoStop}%
\bibitem [{\citenamefont {Wu}\ \emph {et~al.}(2015)\citenamefont {Wu},
  \citenamefont {Buckley}, \citenamefont {Schaibley}, \citenamefont {Feng},
  \citenamefont {Yan}, \citenamefont {Mandrus}, \citenamefont {Hatami},
  \citenamefont {Yao}, \citenamefont {Vu{\v{c}}kovi{\'{c}}}, \citenamefont
  {Majumdar},\ and\ \citenamefont {Xu}}]{Wu2015}%
  \BibitemOpen
  \bibfield  {author} {\bibinfo {author} {\bibfnamefont {S.}~\bibnamefont
  {Wu}}, \bibinfo {author} {\bibfnamefont {S.}~\bibnamefont {Buckley}},
  \bibinfo {author} {\bibfnamefont {J.~R.}\ \bibnamefont {Schaibley}}, \bibinfo
  {author} {\bibfnamefont {L.}~\bibnamefont {Feng}}, \bibinfo {author}
  {\bibfnamefont {J.}~\bibnamefont {Yan}}, \bibinfo {author} {\bibfnamefont
  {D.~G.}\ \bibnamefont {Mandrus}}, \bibinfo {author} {\bibfnamefont
  {F.}~\bibnamefont {Hatami}}, \bibinfo {author} {\bibfnamefont
  {W.}~\bibnamefont {Yao}}, \bibinfo {author} {\bibfnamefont {J.}~\bibnamefont
  {Vu{\v{c}}kovi{\'{c}}}}, \bibinfo {author} {\bibfnamefont {A.}~\bibnamefont
  {Majumdar}},\ and\ \bibinfo {author} {\bibfnamefont {X.}~\bibnamefont {Xu}},\
  }\bibfield  {title} {\bibinfo {title} {Monolayer semiconductor nanocavity
  lasers with ultralow thresholds},\ }\href
  {https://doi.org/10.1038/nature14290} {\bibfield  {journal} {\bibinfo
  {journal} {Nature}\ }\textbf {\bibinfo {volume} {520}},\ \bibinfo {pages}
  {69} (\bibinfo {year} {2015})}\BibitemShut {NoStop}%
\bibitem [{\citenamefont {Fryett}\ \emph {et~al.}(2016)\citenamefont {Fryett},
  \citenamefont {Seyler}, \citenamefont {Zheng}, \citenamefont {Liu},
  \citenamefont {Xu},\ and\ \citenamefont {Majumdar}}]{Fryett2016}%
  \BibitemOpen
  \bibfield  {author} {\bibinfo {author} {\bibfnamefont {T.~K.}\ \bibnamefont
  {Fryett}}, \bibinfo {author} {\bibfnamefont {K.~L.}\ \bibnamefont {Seyler}},
  \bibinfo {author} {\bibfnamefont {J.}~\bibnamefont {Zheng}}, \bibinfo
  {author} {\bibfnamefont {C.-H.}\ \bibnamefont {Liu}}, \bibinfo {author}
  {\bibfnamefont {X.}~\bibnamefont {Xu}},\ and\ \bibinfo {author}
  {\bibfnamefont {A.}~\bibnamefont {Majumdar}},\ }\bibfield  {title} {\bibinfo
  {title} {Silicon photonic crystal cavity enhanced second-harmonic generation
  from monolayer {WS}e$_2$},\ }\href
  {https://doi.org/10.1088/2053-1583/4/1/015031} {\bibfield  {journal}
  {\bibinfo  {journal} {2D Mater.}\ }\textbf {\bibinfo {volume} {4}},\ \bibinfo
  {pages} {015031} (\bibinfo {year} {2016})}\BibitemShut {NoStop}%
\bibitem [{\citenamefont {Fujii}\ and\ \citenamefont
  {Tanabe}(2020)}]{Fujii2020}%
  \BibitemOpen
  \bibfield  {author} {\bibinfo {author} {\bibfnamefont {S.}~\bibnamefont
  {Fujii}}\ and\ \bibinfo {author} {\bibfnamefont {T.}~\bibnamefont {Tanabe}},\
  }\bibfield  {title} {\bibinfo {title} {Dispersion engineering and measurement
  of whispering gallery mode microresonator for {K}err frequency comb
  generation},\ }\href {https://doi.org/doi:10.1515/nanoph-2019-0497}
  {\bibfield  {journal} {\bibinfo  {journal} {Nanophotonics}\ }\textbf
  {\bibinfo {volume} {9}},\ \bibinfo {pages} {1087} (\bibinfo {year}
  {2020})}\BibitemShut {NoStop}%
\bibitem [{\citenamefont {Li}\ \emph {et~al.}(2013{\natexlab{b}})\citenamefont
  {Li}, \citenamefont {Rao}, \citenamefont {Mak}, \citenamefont {You},
  \citenamefont {Wang}, \citenamefont {Dean},\ and\ \citenamefont
  {Heinz}}]{Li2013}%
  \BibitemOpen
  \bibfield  {author} {\bibinfo {author} {\bibfnamefont {Y.}~\bibnamefont
  {Li}}, \bibinfo {author} {\bibfnamefont {Y.}~\bibnamefont {Rao}}, \bibinfo
  {author} {\bibfnamefont {K.~F.}\ \bibnamefont {Mak}}, \bibinfo {author}
  {\bibfnamefont {Y.}~\bibnamefont {You}}, \bibinfo {author} {\bibfnamefont
  {S.}~\bibnamefont {Wang}}, \bibinfo {author} {\bibfnamefont {C.~R.}\
  \bibnamefont {Dean}},\ and\ \bibinfo {author} {\bibfnamefont {T.~F.}\
  \bibnamefont {Heinz}},\ }\bibfield  {title} {\bibinfo {title} {Probing
  symmetry properties of few-layer {M}o{S}$_2$ and h-{BN} by optical
  second-harmonic generation},\ }\href {https://doi.org/10.1021/nl401561r}
  {\bibfield  {journal} {\bibinfo  {journal} {Nano Lett.}\ }\textbf {\bibinfo
  {volume} {13}},\ \bibinfo {pages} {3329} (\bibinfo {year}
  {2013}{\natexlab{b}})}\BibitemShut {NoStop}%
\bibitem [{\citenamefont {Kumar}\ \emph {et~al.}(2013)\citenamefont {Kumar},
  \citenamefont {Najmaei}, \citenamefont {Cui}, \citenamefont {Ceballos},
  \citenamefont {Ajayan}, \citenamefont {Lou},\ and\ \citenamefont
  {Zhao}}]{Kumar2013}%
  \BibitemOpen
  \bibfield  {author} {\bibinfo {author} {\bibfnamefont {N.}~\bibnamefont
  {Kumar}}, \bibinfo {author} {\bibfnamefont {S.}~\bibnamefont {Najmaei}},
  \bibinfo {author} {\bibfnamefont {Q.}~\bibnamefont {Cui}}, \bibinfo {author}
  {\bibfnamefont {F.}~\bibnamefont {Ceballos}}, \bibinfo {author}
  {\bibfnamefont {P.~M.}\ \bibnamefont {Ajayan}}, \bibinfo {author}
  {\bibfnamefont {J.}~\bibnamefont {Lou}},\ and\ \bibinfo {author}
  {\bibfnamefont {H.}~\bibnamefont {Zhao}},\ }\bibfield  {title} {\bibinfo
  {title} {Second harmonic microscopy of monolayer {M}o{S}$_{2}$},\ }\href
  {https://doi.org/10.1103/PhysRevB.87.161403} {\bibfield  {journal} {\bibinfo
  {journal} {Phys. Rev. B}\ }\textbf {\bibinfo {volume} {87}},\ \bibinfo
  {pages} {161403} (\bibinfo {year} {2013})}\BibitemShut {NoStop}%
\bibitem [{\citenamefont {Liu}\ \emph {et~al.}(2016)\citenamefont {Liu},
  \citenamefont {Sinclair}, \citenamefont {Saravi}, \citenamefont {Keeler},
  \citenamefont {Yang}, \citenamefont {Reno}, \citenamefont {Peake},
  \citenamefont {Setzpfandt}, \citenamefont {Staude}, \citenamefont {Pertsch},\
  and\ \citenamefont {Brener}}]{Liu2016}%
  \BibitemOpen
  \bibfield  {author} {\bibinfo {author} {\bibfnamefont {S.}~\bibnamefont
  {Liu}}, \bibinfo {author} {\bibfnamefont {M.~B.}\ \bibnamefont {Sinclair}},
  \bibinfo {author} {\bibfnamefont {S.}~\bibnamefont {Saravi}}, \bibinfo
  {author} {\bibfnamefont {G.~A.}\ \bibnamefont {Keeler}}, \bibinfo {author}
  {\bibfnamefont {Y.}~\bibnamefont {Yang}}, \bibinfo {author} {\bibfnamefont
  {J.}~\bibnamefont {Reno}}, \bibinfo {author} {\bibfnamefont {G.~M.}\
  \bibnamefont {Peake}}, \bibinfo {author} {\bibfnamefont {F.}~\bibnamefont
  {Setzpfandt}}, \bibinfo {author} {\bibfnamefont {I.}~\bibnamefont {Staude}},
  \bibinfo {author} {\bibfnamefont {T.}~\bibnamefont {Pertsch}},\ and\ \bibinfo
  {author} {\bibfnamefont {I.}~\bibnamefont {Brener}},\ }\bibfield  {title}
  {\bibinfo {title} {Resonantly enhanced second-harmonic generation using
  {III-V} semiconductor all-dielectric metasurfaces},\ }\href
  {https://doi.org/10.1021/acs.nanolett.6b01816} {\bibfield  {journal}
  {\bibinfo  {journal} {Nano Lett.}\ }\textbf {\bibinfo {volume} {16}},\
  \bibinfo {pages} {5426} (\bibinfo {year} {2016})}\BibitemShut {NoStop}%
\bibitem [{\citenamefont {Chen}\ \emph {et~al.}(2017)\citenamefont {Chen},
  \citenamefont {Corboliou}, \citenamefont {Solntsev}, \citenamefont {Choi},
  \citenamefont {Vincenti}, \citenamefont {de~Ceglia}, \citenamefont
  {de~Angelis}, \citenamefont {Lu},\ and\ \citenamefont {Neshev}}]{Chen2017}%
  \BibitemOpen
  \bibfield  {author} {\bibinfo {author} {\bibfnamefont {H.}~\bibnamefont
  {Chen}}, \bibinfo {author} {\bibfnamefont {V.}~\bibnamefont {Corboliou}},
  \bibinfo {author} {\bibfnamefont {A.~S.}\ \bibnamefont {Solntsev}}, \bibinfo
  {author} {\bibfnamefont {D.-Y.}\ \bibnamefont {Choi}}, \bibinfo {author}
  {\bibfnamefont {M.~A.}\ \bibnamefont {Vincenti}}, \bibinfo {author}
  {\bibfnamefont {D.}~\bibnamefont {de~Ceglia}}, \bibinfo {author}
  {\bibfnamefont {C.}~\bibnamefont {de~Angelis}}, \bibinfo {author}
  {\bibfnamefont {Y.}~\bibnamefont {Lu}},\ and\ \bibinfo {author}
  {\bibfnamefont {D.~N.}\ \bibnamefont {Neshev}},\ }\bibfield  {title}
  {\bibinfo {title} {Enhanced second-harmonic generation from two-dimensional
  {M}o{S}e$_2$ on a silicon waveguide},\ }\href
  {https://doi.org/10.1038/lsa.2017.60} {\bibfield  {journal} {\bibinfo
  {journal} {Light Sci. Appl.}\ }\textbf {\bibinfo {volume} {6}},\ \bibinfo
  {pages} {e17060} (\bibinfo {year} {2017})}\BibitemShut {NoStop}%
\bibitem [{\citenamefont {Carmon}\ \emph {et~al.}(2004)\citenamefont {Carmon},
  \citenamefont {Yang},\ and\ \citenamefont {Vahala}}]{Carmon2004}%
  \BibitemOpen
  \bibfield  {author} {\bibinfo {author} {\bibfnamefont {T.}~\bibnamefont
  {Carmon}}, \bibinfo {author} {\bibfnamefont {L.}~\bibnamefont {Yang}},\ and\
  \bibinfo {author} {\bibfnamefont {K.~J.}\ \bibnamefont {Vahala}},\ }\bibfield
   {title} {\bibinfo {title} {Dynamical thermal behavior and thermal
  self-stability of microcavities},\ }\href
  {https://doi.org/10.1364/OPEX.12.004742} {\bibfield  {journal} {\bibinfo
  {journal} {Opt. Express}\ }\textbf {\bibinfo {volume} {12}},\ \bibinfo
  {pages} {4742} (\bibinfo {year} {2004})}\BibitemShut {NoStop}%
\bibitem [{\citenamefont {Zhang}\ \emph {et~al.}(2019)\citenamefont {Zhang},
  \citenamefont {Cao}, \citenamefont {Wang}, \citenamefont {Liu}, \citenamefont
  {Qiu}, \citenamefont {Yang}, \citenamefont {Gong},\ and\ \citenamefont
  {Xiao}}]{Zhang2019}%
  \BibitemOpen
  \bibfield  {author} {\bibinfo {author} {\bibfnamefont {X.}~\bibnamefont
  {Zhang}}, \bibinfo {author} {\bibfnamefont {Q.-T.}\ \bibnamefont {Cao}},
  \bibinfo {author} {\bibfnamefont {Z.}~\bibnamefont {Wang}}, \bibinfo {author}
  {\bibfnamefont {Y.-x.}\ \bibnamefont {Liu}}, \bibinfo {author} {\bibfnamefont
  {C.-W.}\ \bibnamefont {Qiu}}, \bibinfo {author} {\bibfnamefont
  {L.}~\bibnamefont {Yang}}, \bibinfo {author} {\bibfnamefont {Q.}~\bibnamefont
  {Gong}},\ and\ \bibinfo {author} {\bibfnamefont {Y.-F.}\ \bibnamefont
  {Xiao}},\ }\bibfield  {title} {\bibinfo {title} {Symmetry-breaking-induced
  nonlinear optics at a microcavity surface},\ }\href
  {https://doi.org/10.1038/s41566-018-0297-y} {\bibfield  {journal} {\bibinfo
  {journal} {Nat. Photon.}\ }\textbf {\bibinfo {volume} {13}},\ \bibinfo
  {pages} {21} (\bibinfo {year} {2019})}\BibitemShut {NoStop}%
\bibitem [{\citenamefont {Humphrey}\ \emph {et~al.}(2007)\citenamefont
  {Humphrey}, \citenamefont {Dale}, \citenamefont {Rosenberger},\ and\
  \citenamefont {Bandy}}]{Humphrey2007}%
  \BibitemOpen
  \bibfield  {author} {\bibinfo {author} {\bibfnamefont {M.~J.}\ \bibnamefont
  {Humphrey}}, \bibinfo {author} {\bibfnamefont {E.}~\bibnamefont {Dale}},
  \bibinfo {author} {\bibfnamefont {A.}~\bibnamefont {Rosenberger}},\ and\
  \bibinfo {author} {\bibfnamefont {D.}~\bibnamefont {Bandy}},\ }\bibfield
  {title} {\bibinfo {title} {Calculation of optimal fiber radius and
  whispering-gallery mode spectra for a fiber-coupled microsphere},\ }\href
  {https://doi.org/https://doi.org/10.1016/j.optcom.2006.10.018} {\bibfield
  {journal} {\bibinfo  {journal} {Opt. Commun.}\ }\textbf {\bibinfo {volume}
  {271}},\ \bibinfo {pages} {124} (\bibinfo {year} {2007})}\BibitemShut
  {NoStop}%
\bibitem [{\citenamefont {Jiang}\ \emph {et~al.}(2017)\citenamefont {Jiang},
  \citenamefont {Shao}, \citenamefont {Zhang}, \citenamefont {Yi},
  \citenamefont {Wiersig}, \citenamefont {Wang}, \citenamefont {Gong},
  \citenamefont {Lončar}, \citenamefont {Yang},\ and\ \citenamefont
  {Xiao}}]{Jiang2017}%
  \BibitemOpen
  \bibfield  {author} {\bibinfo {author} {\bibfnamefont {X.}~\bibnamefont
  {Jiang}}, \bibinfo {author} {\bibfnamefont {L.}~\bibnamefont {Shao}},
  \bibinfo {author} {\bibfnamefont {S.-X.}\ \bibnamefont {Zhang}}, \bibinfo
  {author} {\bibfnamefont {X.}~\bibnamefont {Yi}}, \bibinfo {author}
  {\bibfnamefont {J.}~\bibnamefont {Wiersig}}, \bibinfo {author} {\bibfnamefont
  {L.}~\bibnamefont {Wang}}, \bibinfo {author} {\bibfnamefont {Q.}~\bibnamefont
  {Gong}}, \bibinfo {author} {\bibfnamefont {M.}~\bibnamefont {Lončar}},
  \bibinfo {author} {\bibfnamefont {L.}~\bibnamefont {Yang}},\ and\ \bibinfo
  {author} {\bibfnamefont {Y.-F.}\ \bibnamefont {Xiao}},\ }\bibfield  {title}
  {\bibinfo {title} {Chaos-assisted broadband momentum transformation in
  optical microresonators},\ }\href {https://doi.org/10.1126/science.aao0763}
  {\bibfield  {journal} {\bibinfo  {journal} {Science}\ }\textbf {\bibinfo
  {volume} {358}},\ \bibinfo {pages} {344} (\bibinfo {year}
  {2017})}\BibitemShut {NoStop}%
\bibitem [{\citenamefont {Levy}\ \emph {et~al.}(2011)\citenamefont {Levy},
  \citenamefont {Foster}, \citenamefont {Gaeta},\ and\ \citenamefont
  {Lipson}}]{Levy2011}%
  \BibitemOpen
  \bibfield  {author} {\bibinfo {author} {\bibfnamefont {J.~S.}\ \bibnamefont
  {Levy}}, \bibinfo {author} {\bibfnamefont {M.~A.}\ \bibnamefont {Foster}},
  \bibinfo {author} {\bibfnamefont {A.~L.}\ \bibnamefont {Gaeta}},\ and\
  \bibinfo {author} {\bibfnamefont {M.}~\bibnamefont {Lipson}},\ }\bibfield
  {title} {\bibinfo {title} {Harmonic generation in silicon nitride ring
  resonators},\ }\href {https://doi.org/10.1364/OE.19.011415} {\bibfield
  {journal} {\bibinfo  {journal} {Opt. Express}\ }\textbf {\bibinfo {volume}
  {19}},\ \bibinfo {pages} {11415} (\bibinfo {year} {2011})}\BibitemShut
  {NoStop}%
\bibitem [{\citenamefont {Lu}\ \emph {et~al.}(2021)\citenamefont {Lu},
  \citenamefont {Moille}, \citenamefont {Rao}, \citenamefont {Westly},\ and\
  \citenamefont {Srinivasan}}]{Lu2021}%
  \BibitemOpen
  \bibfield  {author} {\bibinfo {author} {\bibfnamefont {X.}~\bibnamefont
  {Lu}}, \bibinfo {author} {\bibfnamefont {G.}~\bibnamefont {Moille}}, \bibinfo
  {author} {\bibfnamefont {A.}~\bibnamefont {Rao}}, \bibinfo {author}
  {\bibfnamefont {D.~A.}\ \bibnamefont {Westly}},\ and\ \bibinfo {author}
  {\bibfnamefont {K.}~\bibnamefont {Srinivasan}},\ }\bibfield  {title}
  {\bibinfo {title} {Efficient photoinduced second-harmonic generation in
  silicon nitride photonics},\ }\href
  {https://doi.org/10.1038/s41566-020-00708-4} {\bibfield  {journal} {\bibinfo
  {journal} {Nat. Photon.}\ }\textbf {\bibinfo {volume} {15}},\ \bibinfo
  {pages} {131} (\bibinfo {year} {2021})}\BibitemShut {NoStop}%
\bibitem [{\citenamefont {Nitiss}\ \emph {et~al.}(2022)\citenamefont {Nitiss},
  \citenamefont {Hu}, \citenamefont {Stroganov},\ and\ \citenamefont
  {Br{\`e}s}}]{Nitiss2022}%
  \BibitemOpen
  \bibfield  {author} {\bibinfo {author} {\bibfnamefont {E.}~\bibnamefont
  {Nitiss}}, \bibinfo {author} {\bibfnamefont {J.}~\bibnamefont {Hu}}, \bibinfo
  {author} {\bibfnamefont {A.}~\bibnamefont {Stroganov}},\ and\ \bibinfo
  {author} {\bibfnamefont {C.-S.}\ \bibnamefont {Br{\`e}s}},\ }\bibfield
  {title} {\bibinfo {title} {Optically reconfigurable quasi-phase-matching in
  silicon nitride microresonators},\ }\href
  {https://doi.org/10.1038/s41566-021-00925-5} {\bibfield  {journal} {\bibinfo
  {journal} {Nat. Photon.}\ }\textbf {\bibinfo {volume} {16}},\ \bibinfo
  {pages} {134} (\bibinfo {year} {2022})}\BibitemShut {NoStop}%
\bibitem [{\citenamefont {Fujii}\ \emph {et~al.}(2023)\citenamefont {Fujii},
  \citenamefont {Wada}, \citenamefont {Sugano}, \citenamefont {Kumazaki},
  \citenamefont {Kogure}, \citenamefont {Kato},\ and\ \citenamefont
  {Tanabe}}]{Fujii2023}%
  \BibitemOpen
  \bibfield  {author} {\bibinfo {author} {\bibfnamefont {S.}~\bibnamefont
  {Fujii}}, \bibinfo {author} {\bibfnamefont {K.}~\bibnamefont {Wada}},
  \bibinfo {author} {\bibfnamefont {R.}~\bibnamefont {Sugano}}, \bibinfo
  {author} {\bibfnamefont {H.}~\bibnamefont {Kumazaki}}, \bibinfo {author}
  {\bibfnamefont {S.}~\bibnamefont {Kogure}}, \bibinfo {author} {\bibfnamefont
  {Y.~K.}\ \bibnamefont {Kato}},\ and\ \bibinfo {author} {\bibfnamefont
  {T.}~\bibnamefont {Tanabe}},\ }\bibfield  {title} {\bibinfo {title}
  {Versatile tuning of kerr soliton microcombs in crystalline
  microresonators},\ }\href {https://doi.org/10.1038/s42005-022-01118-4}
  {\bibfield  {journal} {\bibinfo  {journal} {Commun. Phys.}\ }\textbf
  {\bibinfo {volume} {6}},\ \bibinfo {pages} {1} (\bibinfo {year}
  {2023})}\BibitemShut {NoStop}%
\end{thebibliography}%
\end{document}